\documentclass[prd,preprint,nofootinbib]{revtex4-1}

\usepackage{amssymb,amsmath}
\usepackage{color}
\usepackage{graphicx,subfigure}
\usepackage{hyperref}
\usepackage{float} 
\usepackage[T1]{fontenc}

\newcommand{\nn}{\nonumber}

\newcommand \vect[1]{
        \left(\begin{matrix} #1 \end{matrix}\right)
}

\newcommand \ket[1]{
        \left| #1 \right>
}

\newcommand \bra[1]{
        \left< #1 \right|
}

\begin{document}

\title{U-Spin Sum Rules for CP Asymmetries of\\ Three-Body Charmed Baryon Decays}

\author{Yuval Grossman}
\email{yg73@cornell.edu}
\affiliation{Department of Physics, LEPP, Cornell University, Ithaca, NY 14853, USA}
\author{Stefan Schacht}
\email{ss3843@cornell.edu}
\affiliation{Department of Physics, LEPP, Cornell University, Ithaca, NY 14853, USA}

\begin{abstract}
Triggered by a recent LHCb measurement and prospects for Belle~II, we derive U-spin symmetry 
relations between integrated CP asymmetries of three-body $\Lambda_c^+$ and $\Xi_c^+$ decays. 
The sum rules read 
$A_{CP}(\Lambda_c^+\rightarrow p K^- K^+) + A_{CP}(\Xi_c^+\rightarrow \Sigma^+ \pi^- \pi^+) = 0$\,,
$A_{CP}(\Lambda_c^+\rightarrow p \pi^- \pi^+) + A_{CP}(\Xi_c^+\rightarrow \Sigma^+ K^- K^+) = 0$\,, and
$A_{CP}(\Lambda_c^+\rightarrow \Sigma^+ \pi^- K^+) + A_{CP}(\Xi_c^+\rightarrow pK^-\pi^+)   = 0$\,.
No such U-spin sum rule exists between $A_{CP}(\Lambda_c^+\rightarrow p K^- K^+)$ and $A_{CP}(\Lambda_c^+\rightarrow p \pi^- \pi^+)$.
All of these sum rules are associated with a complete interchange of $d$ and $s$ quarks. 
Furthermore, there are no U-spin CP asymmetry sum rules which hold to first order U-spin breaking.
\end{abstract}

\maketitle

\section{Introduction}\label{sec:intro}

Recently, after establishing the first evidence for CP violation in beauty baryon decays~\cite{Aaij:2016cla}, 
LHCb measured the difference of CP asymmetries of the three-body singly Cabibbo-suppressed (SCS) $\Lambda_c^+$ decays \cite{Aaij:2017xva}
\begin{align}
A_{CP}(\Lambda_c^+\rightarrow p K^-K^+) - A_{CP}(\Lambda_c^+\rightarrow p \pi^-\pi^+) 
&= (0.30\pm 0.91 \pm 0.61)\%\,. \label{eq:LHCb-measurement}
\end{align}
Here, $A_{CP}$ is the CP asymmetry of the rates integrated over the whole phase space, 
for details see Ref.~\cite{Aaij:2017xva}, we give a formal definition in Eq.~(\ref{eq:phase-space-integrated-ACP}).
Prospects for future improvements are bright~\cite{Bediaga:2018lhg} and there is also a 
rich physics program with charmed baryons at Belle~II~\cite{Schwartz:2018avh,Kou:2018nap}.

For charmed meson decays, sum rules between direct CP asymmetries are known. In the U-spin limit we have for the direct CP 
asymmetries 
(see, e.g. Refs.~\cite{Grossman:2006jg, Pirtskhalava:2011va, Hiller:2012xm, Grossman:2012ry})
\begin{align}
a_{CP}^{\mathrm{dir}}(D^0\rightarrow K^+K^-) + a_{CP}^{\mathrm{dir}}(D^0\rightarrow \pi^+\pi^-) &= 0\,, \label{eq:meson-sum-rule-1} \\
a_{CP}^{\mathrm{dir}}(D^+\rightarrow K_SK^+) + a_{CP}^{\mathrm{dir}}(D_s^+\rightarrow K_S\pi^+) &= 0\,. \label{eq:meson-sum-rule-2}
\end{align}
Generalizations including SU(3)$_F$ breaking effects have also been discussed in the 
literature~\cite{Grossman:2012ry,Grossman:2013lya,Muller:2015rna}.
Consequently, in view of the measurement, Eq.~(\ref{eq:LHCb-measurement}),
the question arises if similar U-spin symmetry relations exist also between the decays involved therein. 
In this letter we address this question. 
We focus therefore only on SCS three-body charmed baryon decays which are related to 
$\Lambda_c^+\rightarrow p K^-K^+$ and $\Lambda_c^+\rightarrow p \pi^-\pi^+$ by U-spin.
These are the decays 
$\Xi_c^+\rightarrow \Sigma^+ \pi^-\pi^+$,  
$\Lambda_c^+\rightarrow \Sigma^+\pi^-K^+$,
$\Xi_c^+\rightarrow p K^-\pi^+$,
and $\Xi_c^+\rightarrow \Sigma^+ K^-K^+$, 
\emph{i.e.},~altogether six decay channels connected by U-spin.

Naively, one could expect that replacing the $D^0$ by a $\Lambda_c^+$ and adding a proton in the 
final states in Eq.~(\ref{eq:meson-sum-rule-1})
would also give a valid sum rule. As we show, however, the presence of the spectator quark has nontrivial implications
as the three-body decay allows more combinatorial possibilities for the 
flavor-flow diagrams. 
The $d$ spectator quark can end in the proton or the pion, but not in the kaon. 
Therefore, it turns out that $\Lambda_c^+\rightarrow p \pi^-\pi^+$ has additional independent topological diagrams which are 
not present in case of $\Lambda_c^+\rightarrow p K^-K^+$ and there is no U-spin sum rule between the two respective CP 
asymmetries. However, we find that analogs of Eq.~(\ref{eq:meson-sum-rule-1}) still exist and correlate 
$\Lambda_c^+$ and $\Xi_c^+$ decays. These sum rules share with Eqs.~(\ref{eq:meson-sum-rule-1}) and (\ref{eq:meson-sum-rule-2}) 
the feature that they come from interchanging all $d$ and $s$ quarks of a given process \cite{Gronau:2000zy,Fleischer:1999pa,Gronau:2000md}.

The symmetries of charm decay amplitudes which lead to correlations between different CP asymmetries
can be expressed in form of topological diagrams or reduced matrix elements from group theory.
After reviewing the available literature on charmed baryon decays in Sec.~\ref{sec:literature}, we introduce 
both parametrizations in Sec.~\ref{sec:parametrizations}. We show that both approaches result in equivalent decompositions.
In Sec.~\ref{sec:sumrules} we discuss how the pointwise CP asymmetries are connected to the integrated ones and  
conclude in Sec.~\ref{sec:conclusions}. 
In the appendix we give the U-spin breaking contributions which show
that no CP asymmetry sum rules exist at first order U-spin breaking.

\section{Literature Review \label{sec:literature}}

A variety of methods has been applied to charm baryon decays in the literature.
Most promising are SU(3)$_F$ methods, however large corrections of $\mathcal{O}(30\%)$ are expected 
from SU(3)$_F$ breaking.
Those symmetry-based methods have been used for two-body charmed baryon decays since a long time, including discussions of 
CP violation~\cite{Altarelli:1975ye, Kingsley:1975fe, Korner:1978tc,  Matsuda:1978cj, Savage:1989qr, Pakvasa:1990if, Savage:1991wt, Verma:1995dk, Sharma:1996sc}, for general reviews see Refs.~\cite{Korner:1991ap, Korner:1994nh, Korner:1996dg, Cheng:2015iom}.
The connection with the diagrammatic approach for two-body decays~\cite{Kohara:1991ug, Chau:1995gk, Kohara:1997nu, Fayyazuddin:1996iy}
and SU(3)$_F$ breaking~\cite{Savage:1991wu} has also been discussed, and even SU(4)$_F$ has been 
applied~\cite{Sheikholeslami:1991ab, Khanna:1993xv}.
More recent works, which however do not discuss CP violation are 
Refs.~\cite{Geng:2017esc,Geng:2017mxn, Jiang:2018iqa, Geng:2018bow}. 
Besides SU(3)$_F$ there have also been several other approaches to two-body charmed baryon decays, like (covariant) quark 
models \cite{Korner:1992wi, Zenczykowski:1993hw, Uppal:1994pt, Gutsche:2018utw}, 
pole and factorization models \cite{Cheng:1991sn, Sharma:2009zzj,Sharma:2009zze}, Heavy Quark 
Effective Theory (HQET) \cite{Korner:1994nh, Sharma:1998rd} and a light-front approach \cite{Zhao:2018zcb}. A comparison of 
several model-dependent approaches is provided in Ref.~\cite{Cheng:2018hwl}. 
The CP violating effect from the interference of charm and neutral kaon decays in two-body charmed baryon decays 
has been discussed in Ref.~\cite{Wang:2017gxe}. CP violation in $\Lambda_c\rightarrow BP$ and $\Lambda_c\rightarrow BV$
(where $B$ is a baryon, $P$ a pseudoscalar meson and $V$ a vector meson) has been discussed in Ref.~\cite{Kang:2010td}, 
however not in the context of SU(3)$_F$ sum rules. 
Prospects for decay asymmetry parameter measurements at BESIII are given in Ref.~\cite{Wang:2016elx}.
There is also literature on using SU(3)$_F$ for decays of baryons with more than 
one charm quark, see Refs.~\cite{Savage:1990pr, Egolf:2002nk, Richard:2002nn, Wang:2017azm, Shi:2017dto, Wang:2018utj}.

Three-body charmed baryon decays have been covered in the SU(3)$_F$ approach 
in Refs.~\cite{Savage:1989qr, Lu:2016ogy, Geng:2018plk, Geng:2018upx}. 
A general analysis of the New Physics (NP) sensitivity of different baryonic decay channels can be found in Ref.~\cite{Bigi:2012ev}.
In Ref.~\cite{Gronau:2018vei} a statistical isospin model has been applied.
However, the CKM-subleading parts which are essential for CP asymmetries are not studied in these references.
Moreover, we were unable to find sum rules for CP asymmetries of three-body charmed baryon decays in the literature,
and this is what we do next.

\section{U-spin decomposition \label{sec:parametrizations}}

In this paper we consider only the Standard Model (SM). Then, the Cabibbo-Kobayashi-Maskawa (CKM) structure of amplitudes of SCS charm decays can be written as
\begin{align}
\mathcal{A} &= \Sigma ( A_{\Sigma}^s - A_{\Sigma}^d) + \Delta A_{\Delta}\,, \label{eq:CKM-structure}
\end{align}
where $A_{\Sigma}^s$, $A_{\Sigma}^d$ and $A_{\Delta}$ carry a strong phase only.
The CKM matrix elements appear in the combinations
\begin{align}
\Sigma &\equiv \frac{V_{cs}^* V_{us} - V_{cd}^* V_{ud}}{2}\,, &
\Delta &\equiv \frac{V_{cs}^* V_{us} + V_{cd}^* V_{ud}}{2} = -\frac{V_{cb}^*V_{ub}}{2}\,, \label{eq:CKMstructure} 
\end{align}
where we used CKM unitarity for $\Delta$.

$A_{\Sigma}^s$ ($A_{\Sigma}^d$) contains $c\rightarrow s$ ($c\rightarrow d$) quark-level transitions. 
Note that for some decays both $A_{\Sigma}^s$ and $A_{\Sigma}^d$ are nonzero, see Table~\ref{tab:quarklevel}.
We have $\Delta \ll \Sigma$, thus, $A_{\Sigma} \equiv A_{\Sigma}^s - A_{\Sigma}^d$ is the CKM-leading part, 
whereas $A_{\Delta}$ is CKM-subleading.
Actually, the contribution of $\Delta A_{\Delta}$ is negligible 
for the current and near-future experimental precision of branching ratio measurements. 
However, the interference of $\Delta A_{\Delta}$ with $\Sigma A_{\Sigma}$ is essential for non-vanishing direct charm CP asymmetries.

For deriving the diagrammatic and group-theoretical parametrizations we use the following conventions 
for the quark flavor states of the relevant baryon and meson states, which are compatible with 
Refs.~\cite{Gronau:1994rj, Gronau:1995hm, Soni:2006vi, Muller:2015lua}
\begin{align}
\left|\Lambda_c^+\right> &\equiv \ket{udc} = \ket{\frac{1}{2},\frac{1}{2}}  \,,  &  
\ket{\Xi_c^+}  	  &\equiv \ket{usc} = \ket{\frac{1}{2},-\frac{1}{2}} \,,\nn\\ 
\ket{p} 	  &\equiv \ket{uud} = \ket{\frac{1}{2},\frac{1}{2}}  \,,&
\ket{\Sigma^+} 	  &\equiv \ket{uus} = \ket{\frac{1}{2},-\frac{1}{2}}  \,,\nn\\
\ket{K^+}	  &\equiv \ket{u\bar{s}} =  \ket{\frac{1}{2},\frac{1}{2}}   \,,&
\ket{\pi^+}	  &\equiv \ket{u\bar{d}} = -\ket{\frac{1}{2},-\frac{1}{2}}  \,,\nn\\
-\ket{\pi^-}  	  &\equiv \ket{d\bar{u}} = \ket{\frac{1}{2},\frac{1}{2}}    \,,&
-\ket{K^-}	  &\equiv \ket{s\bar{u}} = \ket{\frac{1}{2},-\frac{1}{2}}   \,,\nn
\end{align}
and where we write the states as U-spin doublets.
The operators of the effective Hamiltonian for SCS decays can be written as
a sum of spurions with $\Delta U=1$ and $\Delta U=0$: 
\begin{align}
\mathcal{H}_{\mathrm{eff}} \sim \Sigma \, (1,0)  + \Delta \,(0,0) \,, 
\end{align}
where $(i,j) \equiv \mathcal{O}^{\Delta U=i}_{\Delta U_3=j}$, see the discussion in Ref.~\cite{Brod:2012ud}. We show here the flavor structure 
with respect to U-spin only, absorbing any overall factors into the group representations.

The group-theoretical decomposition is obtained by applying the Wigner-Eckart theorem. 
For the final states, we use the order $(B\otimes P^-)\otimes P^+$,~\emph{i.e.},~we calculate first the 
tensor product of the baryon 
with the negatively charged pseudoscalar and then we calculate the tensor product of the result with the positively 
charged pseudoscalar.
For the final state $\bra{\frac{1}{2}}$ we put a subscript \lq\lq{}$0$\rq\rq{} or \lq\lq{}$1$\rq\rq{} depending on 
whether it comes from the tensor product $0\times\frac{1}{2}$ or $1\times\frac{1}{2}$, respectively, and we distinguish the
corresponding reduced matrix elements. Our result is shown in Table~\ref{tab:group}.
 
\begin{figure}[h]
\begin{center}
\subfigure[\,$T$]{\includegraphics[width=0.23\textwidth]{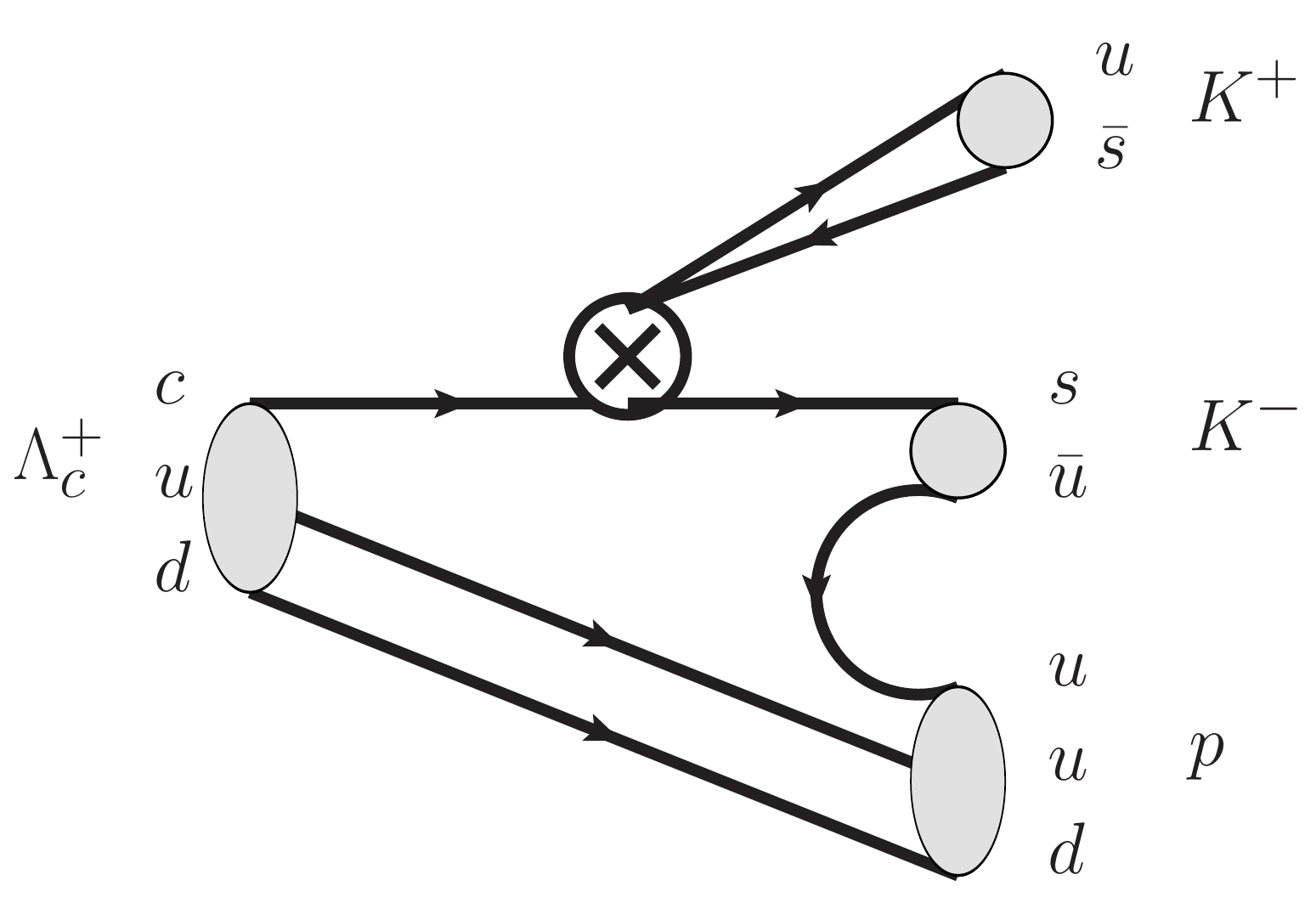}}
\subfigure[\,$C_1$]{\includegraphics[width=0.23\textwidth]{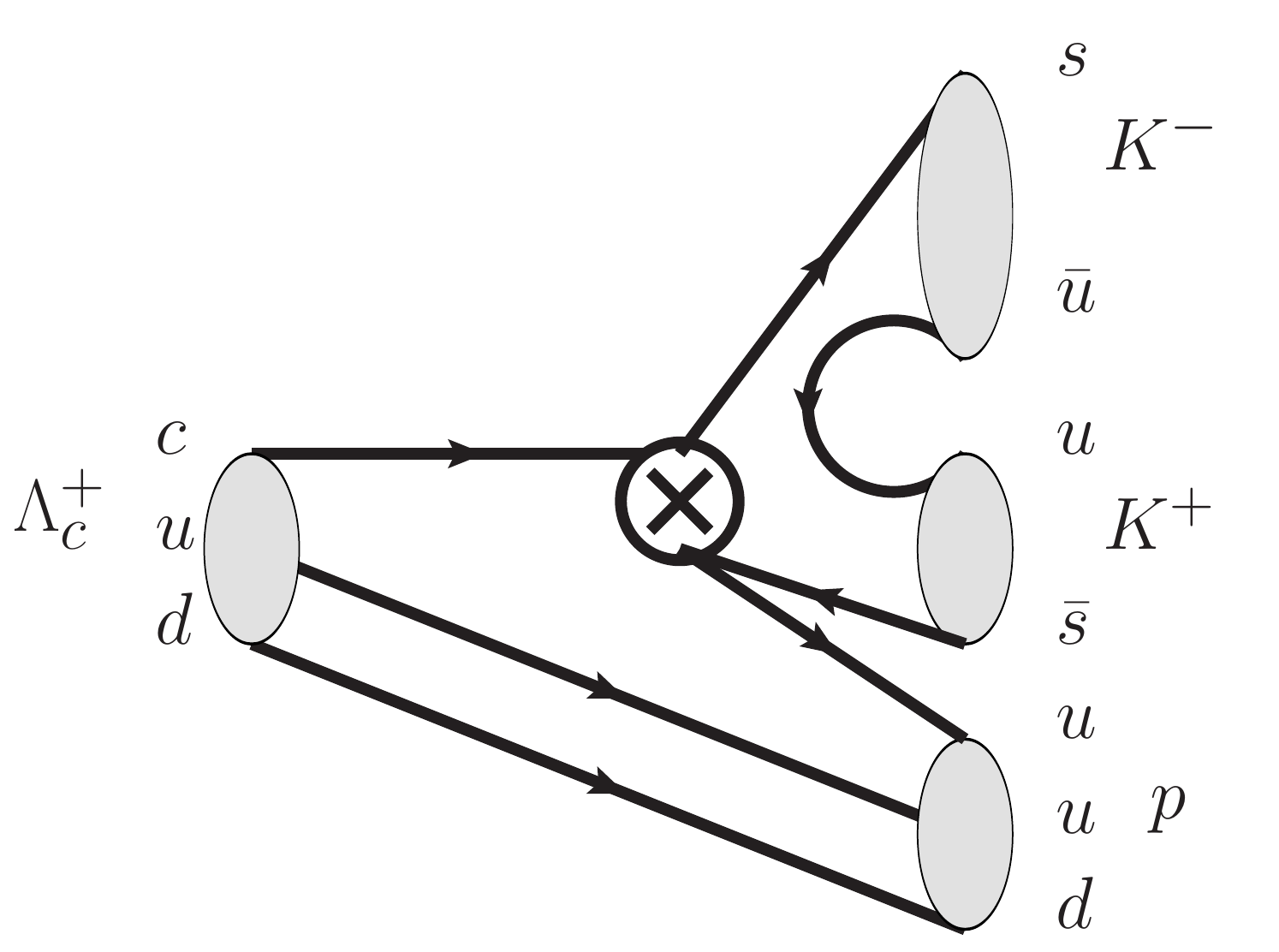}} 
\subfigure[\,$E_2$]{\includegraphics[width=0.23\textwidth]{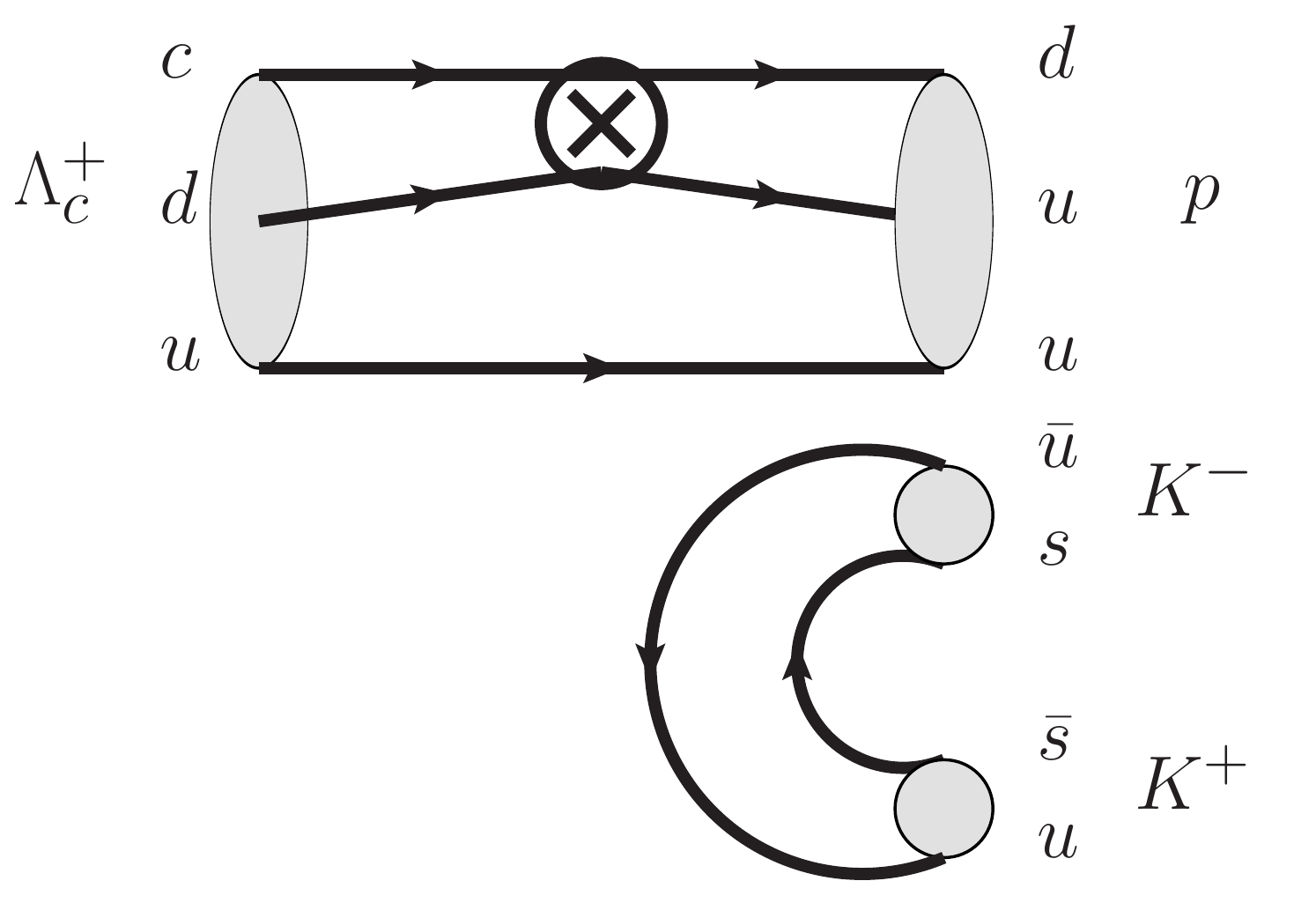}} 
\subfigure[\,$P_1$]{\includegraphics[width=0.23\textwidth]{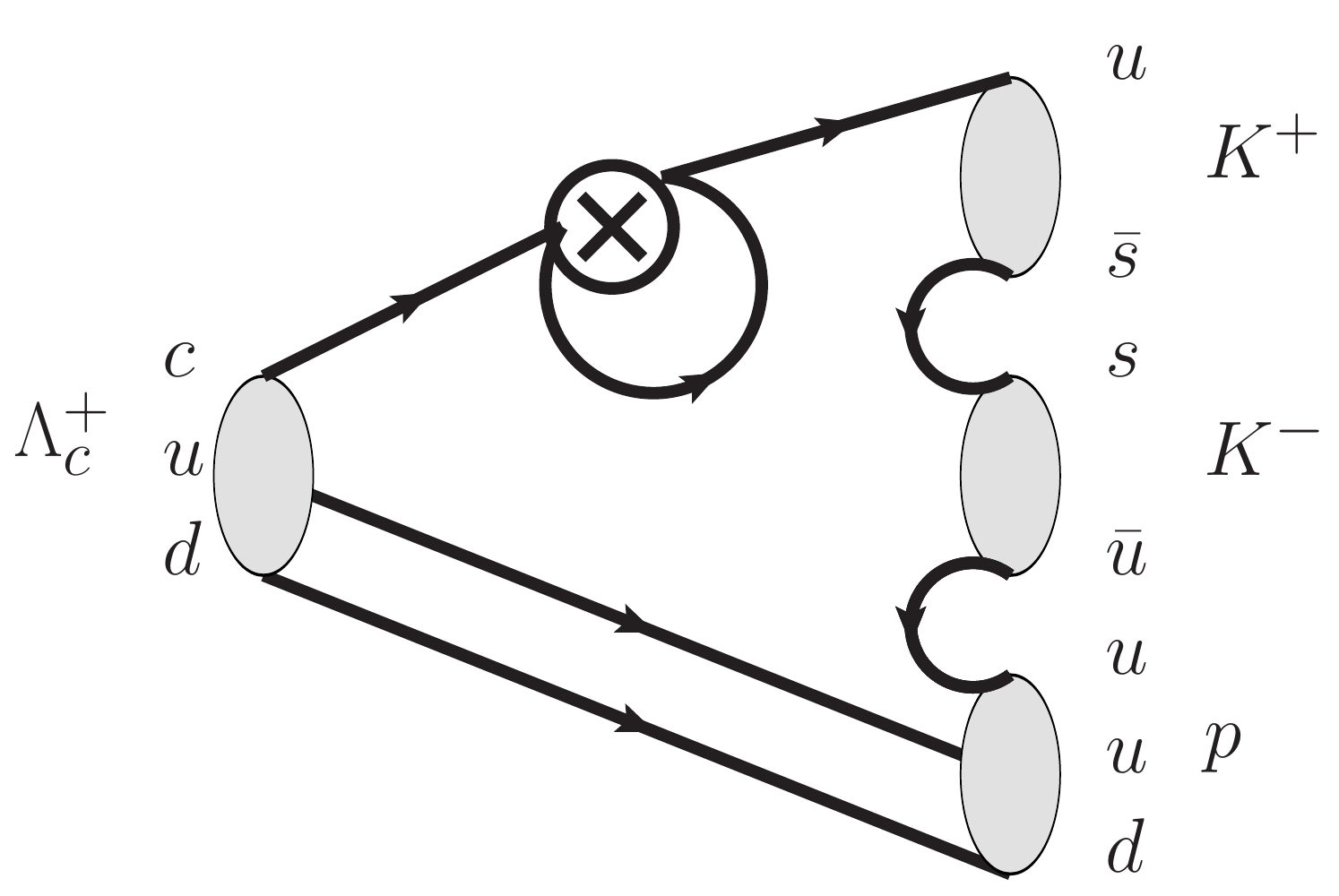}}
\caption{Diagrams in $\mathcal{A}(\Lambda_c^+\rightarrow pK^-K^+) = (\Sigma+\Delta) (-T - C_1) + (-\Sigma+\Delta) (-E_2) +\Delta (-P_1)$.
All diagrams have been drawn using \texttt{JaxoDraw}~\cite{Binosi:2003yf,Vermaseren:1994je}.
}
\label{fig:Lambdac-pKK}
\end{center}
\end{figure}

\begin{figure}[h]
\begin{center}
\subfigure[\,$T$]{\includegraphics[width=0.23\textwidth]{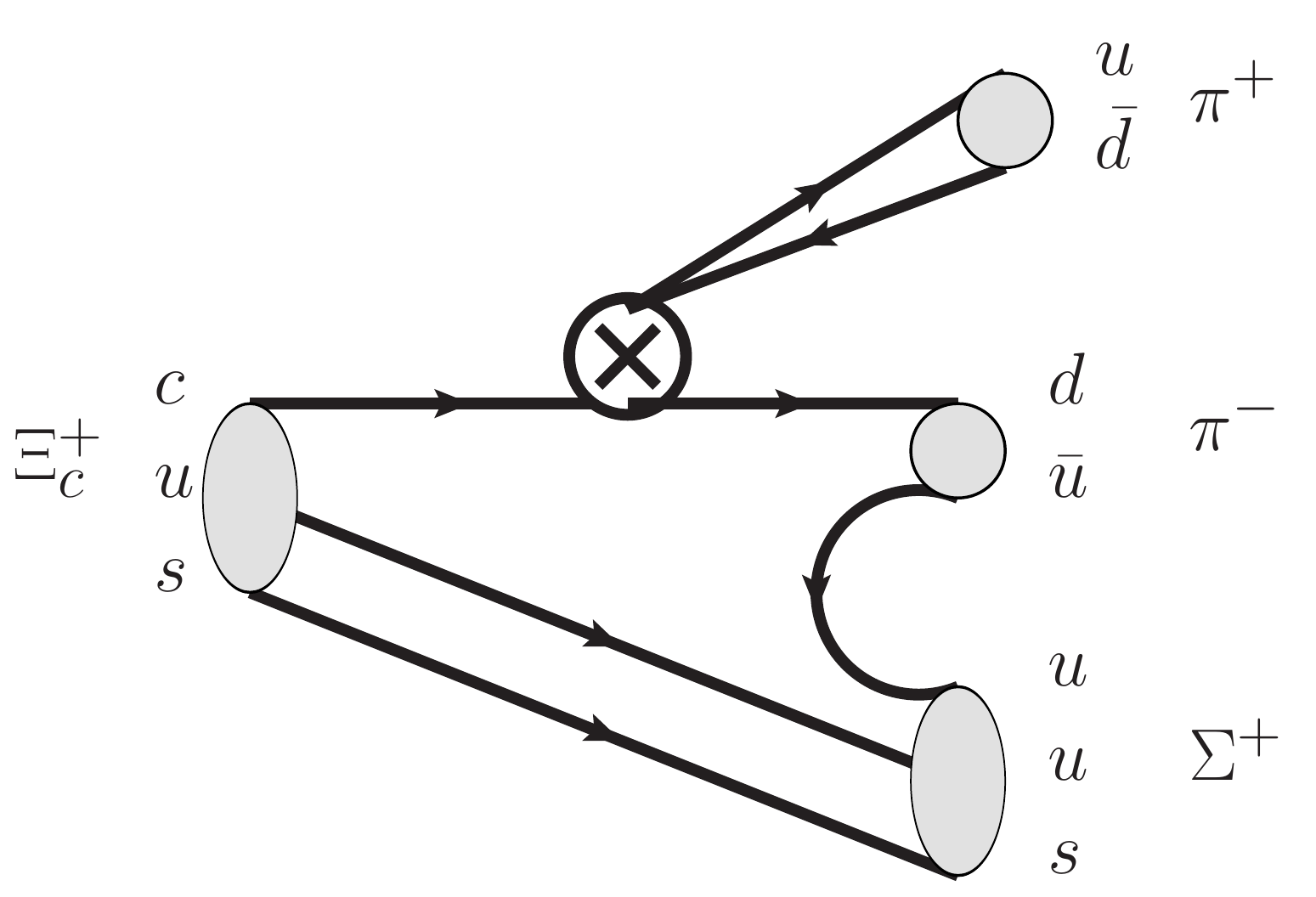}}
\subfigure[\,$C_1$]{\includegraphics[width=0.23\textwidth]{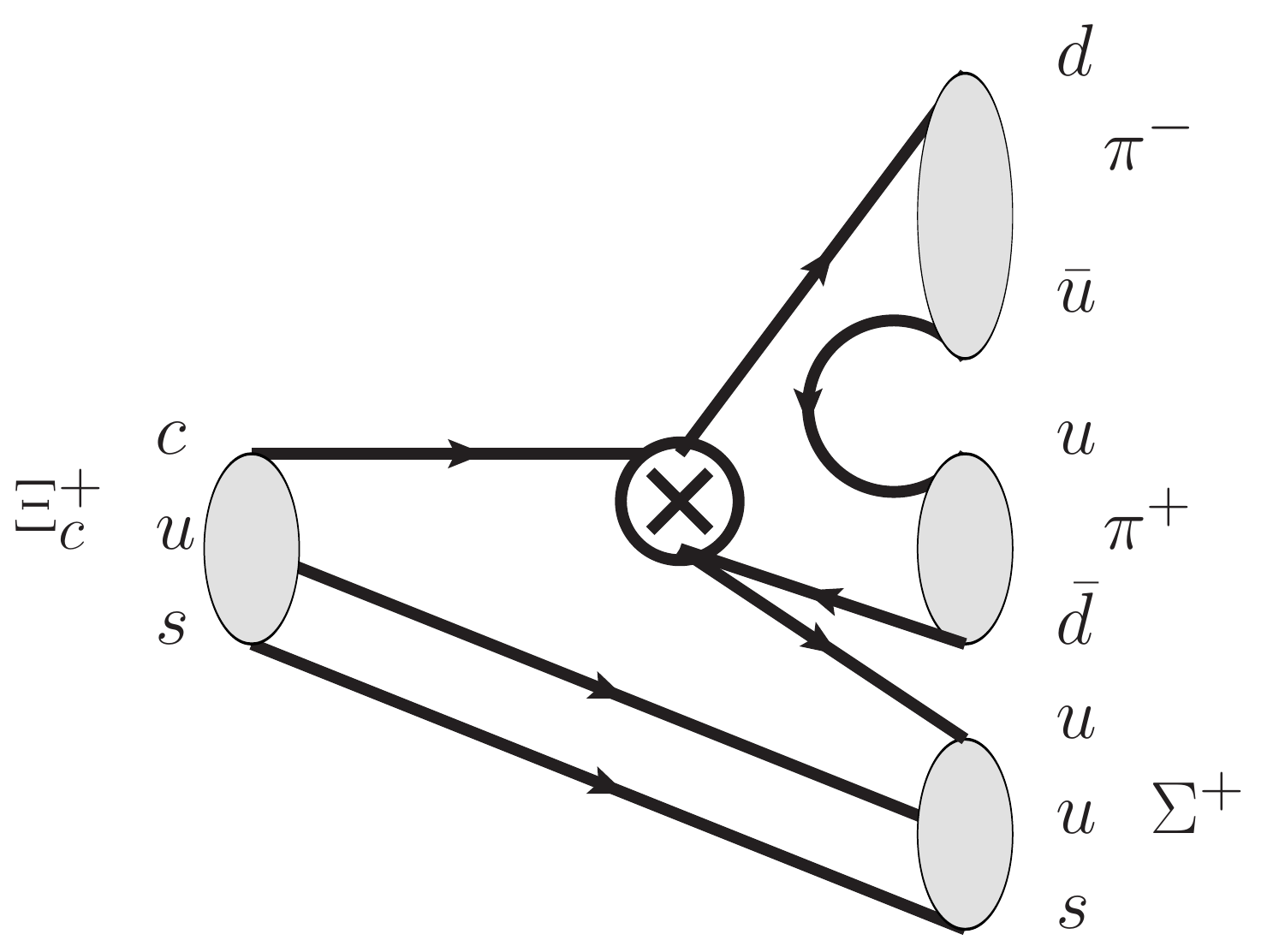}} 
\subfigure[\,$E_2$]{\includegraphics[width=0.23\textwidth]{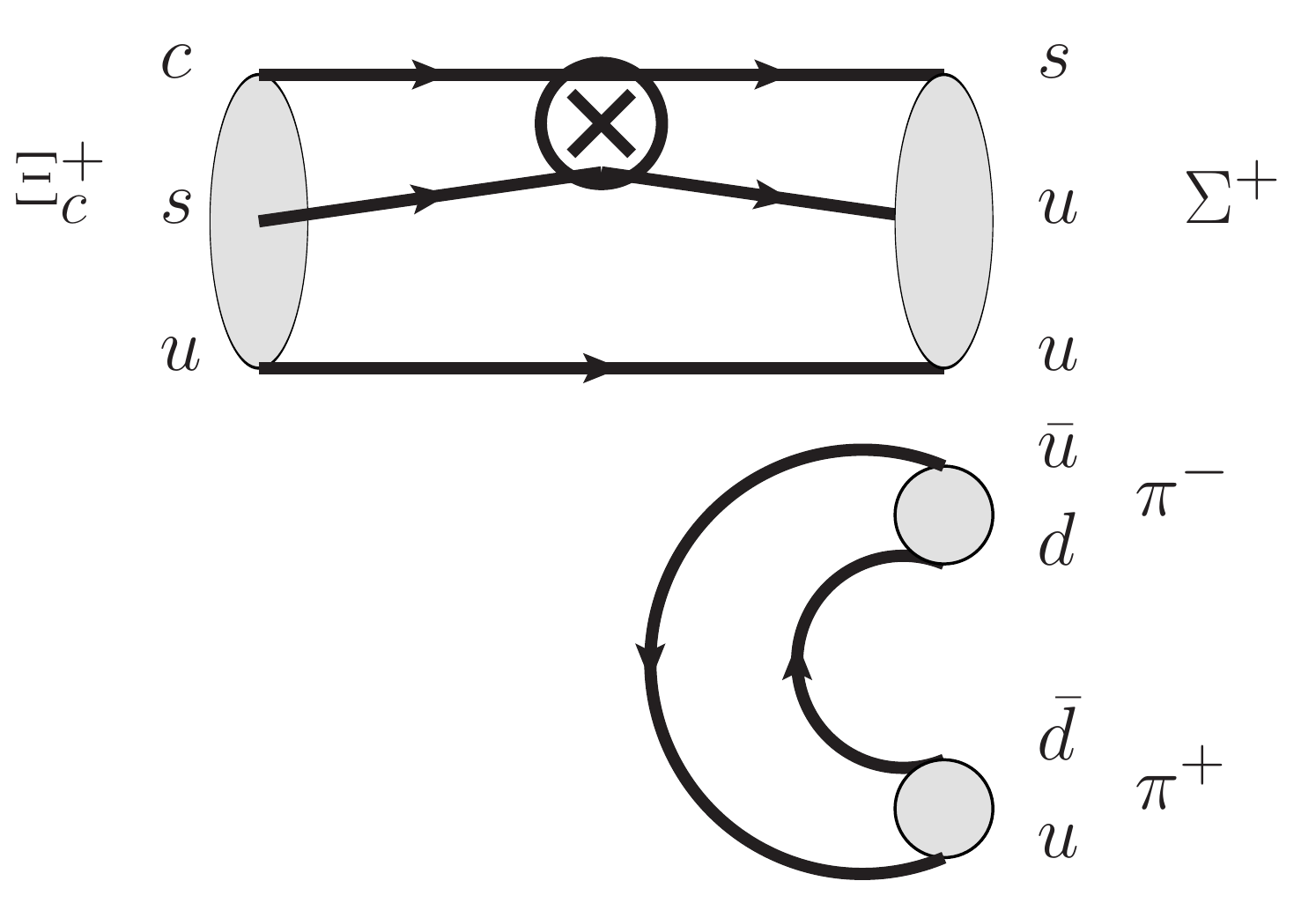}} 
\subfigure[\,$P_1$]{\includegraphics[width=0.23\textwidth]{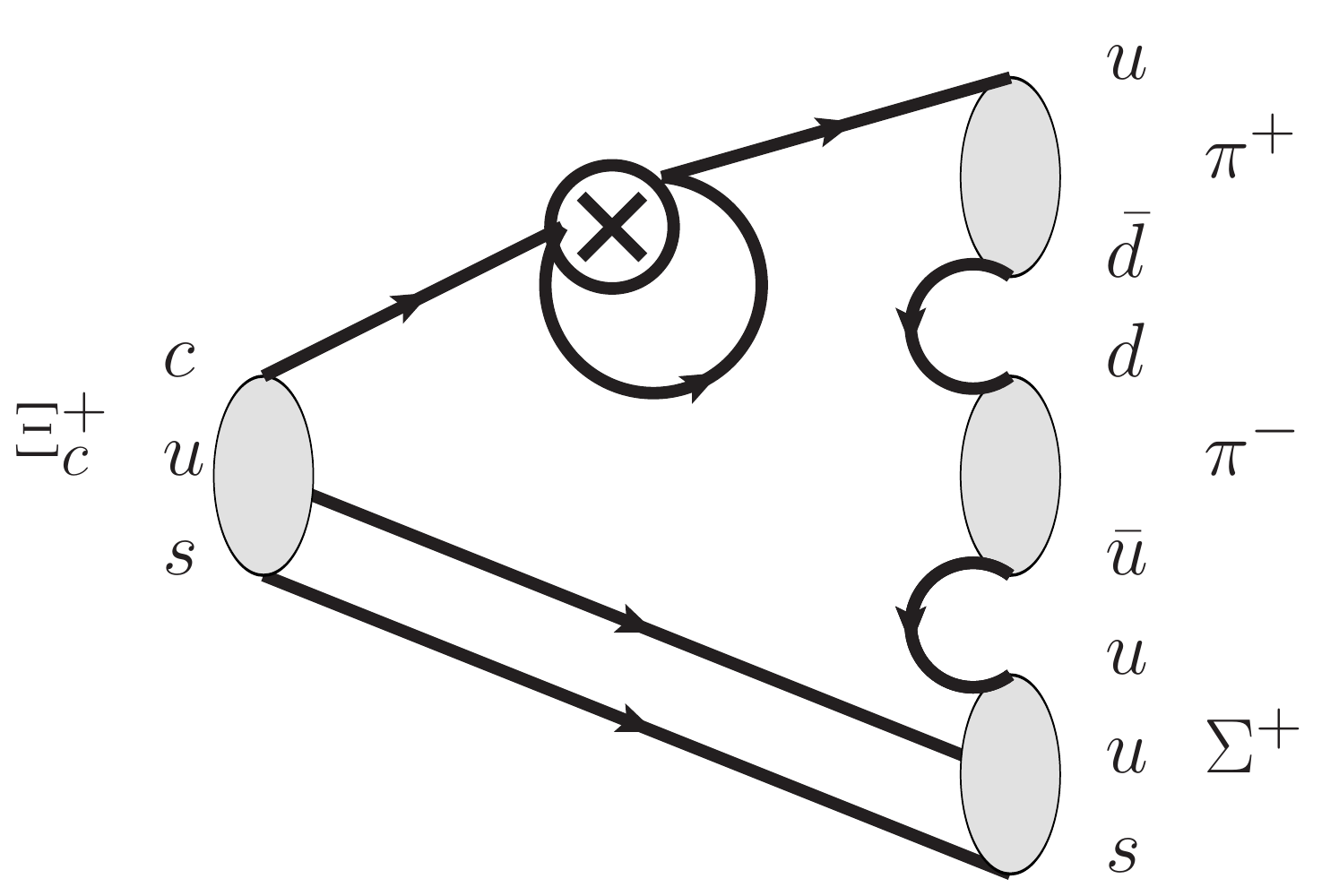}}
\caption{Diagrams in $\mathcal{A}(\Xi_c^+\rightarrow \Sigma^+\pi^-\pi^+) = (-\Sigma+\Delta) (-T - C_1) + (\Sigma+\Delta) (-E_2) +\Delta (-P_1)$. 
}
\label{fig:Xic-SigmaPiPi}
\end{center}
\end{figure}

\begin{figure}[h]
\begin{center}
\subfigure[\,$C_2$]{\includegraphics[width=0.23\textwidth]{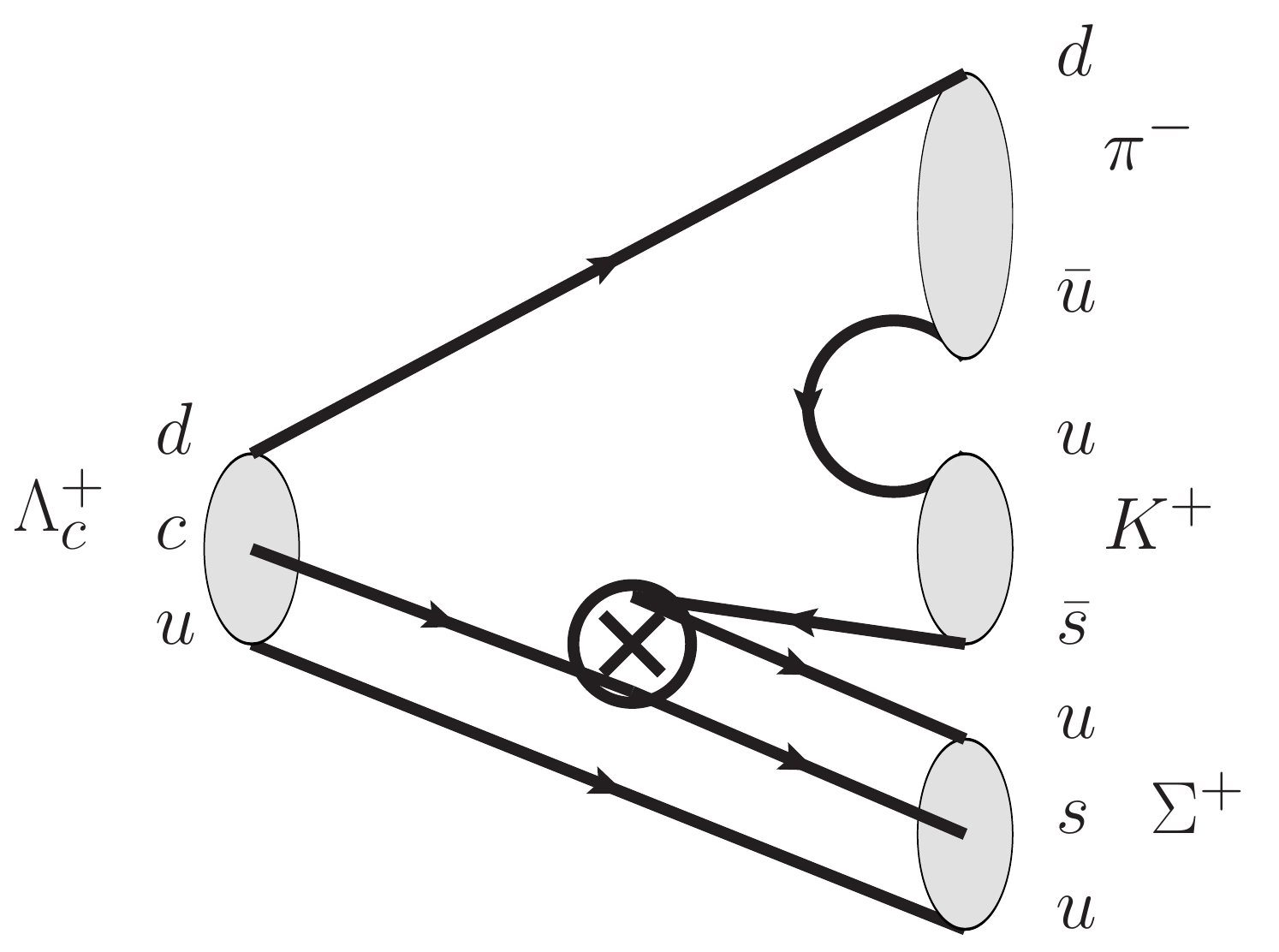}}
\subfigure[\,$E_1$]{\includegraphics[width=0.23\textwidth]{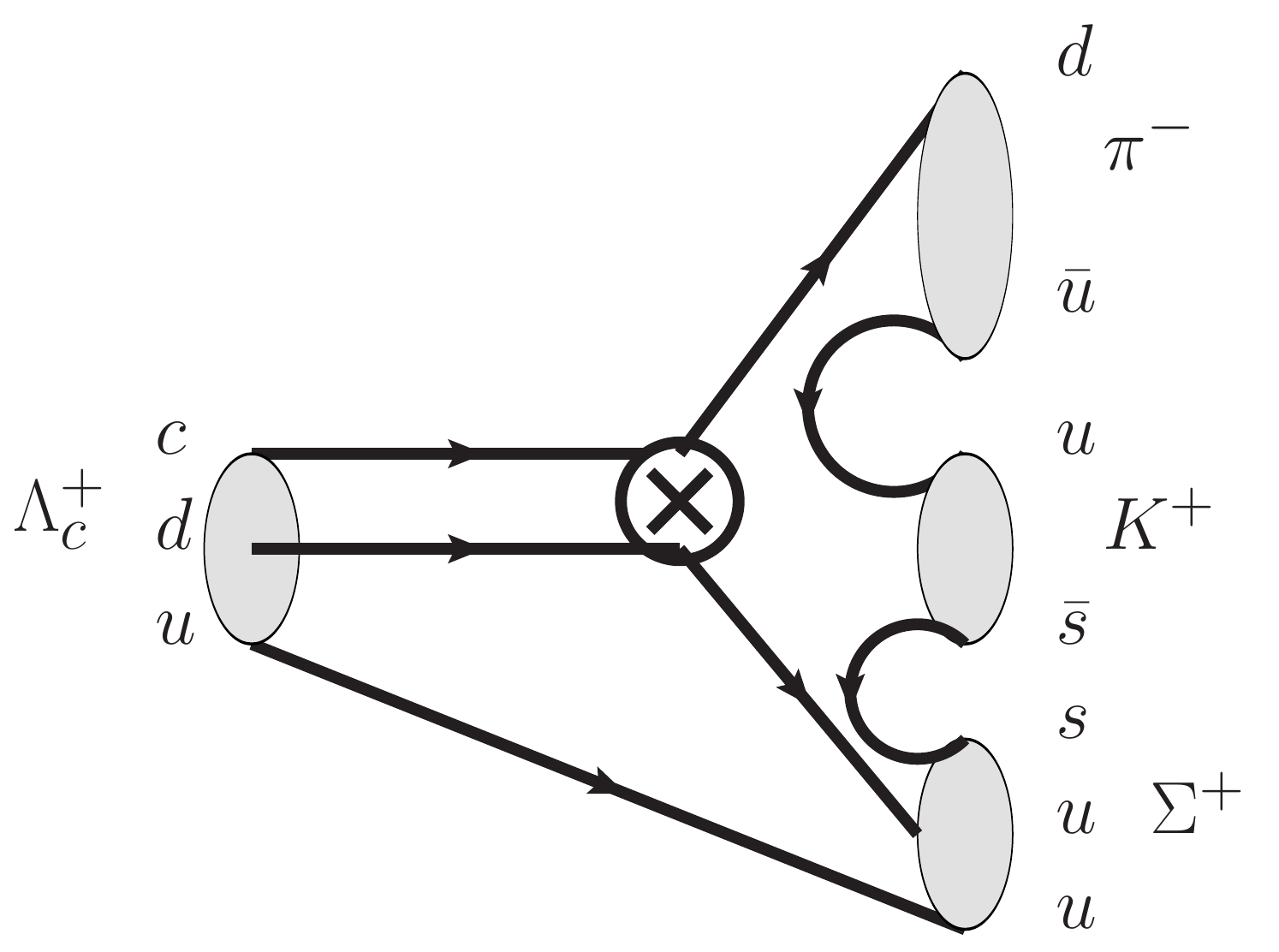}} 
\subfigure[\,$P_2$]{\includegraphics[width=0.23\textwidth]{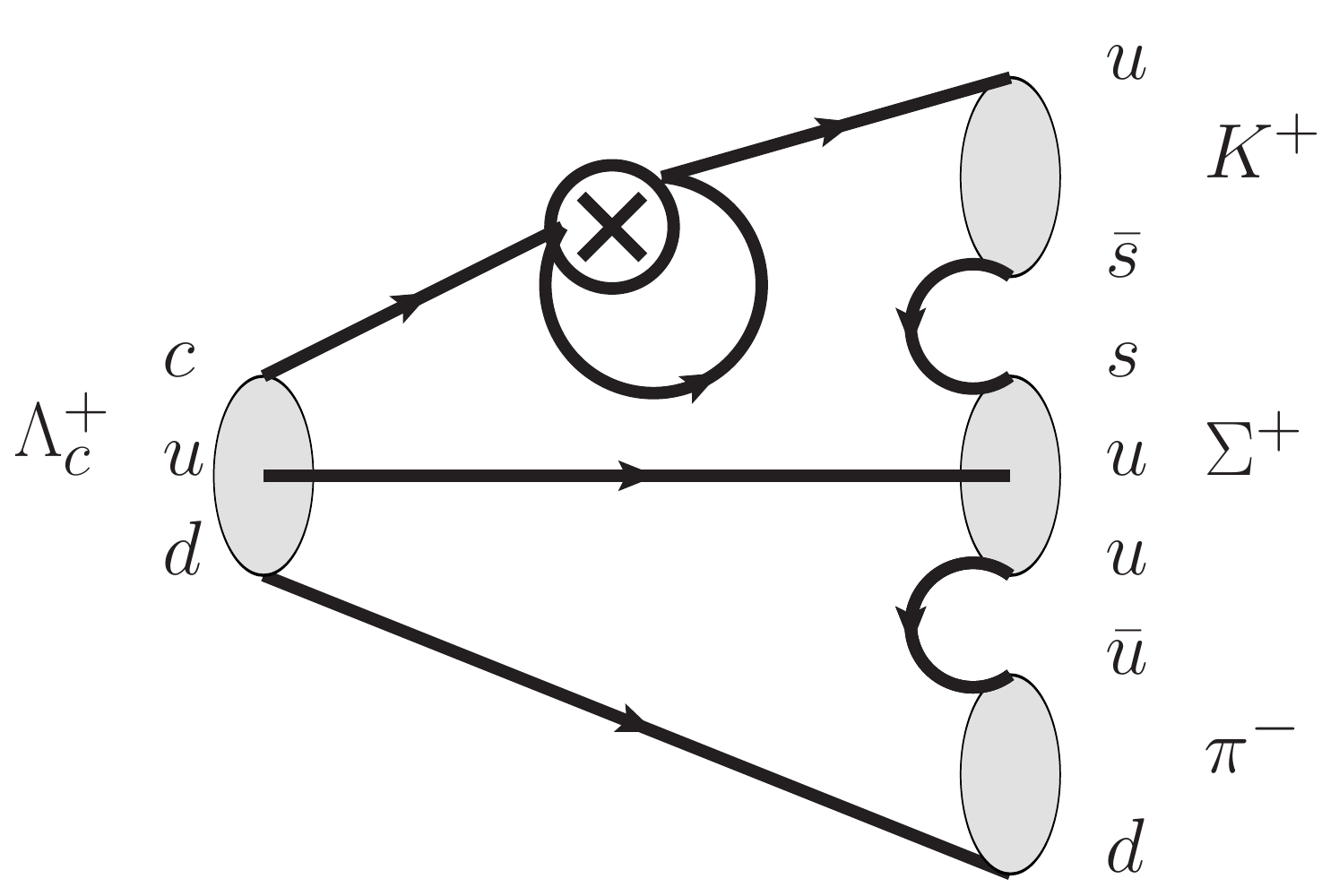}}
\caption{Diagrams in $\mathcal{A}(\Lambda_c^+\rightarrow \Sigma^+\pi^-K^+) = (\Sigma+\Delta) (-C_2) + (-\Sigma+\Delta) (-E_1) +\Delta (-P_2)$. 
}
\label{fig:Lambdac-SigmaPiK}
\end{center}
\end{figure}

\begin{figure}[h]
\begin{center}
\subfigure[\,$C_2$]{\includegraphics[width=0.23\textwidth]{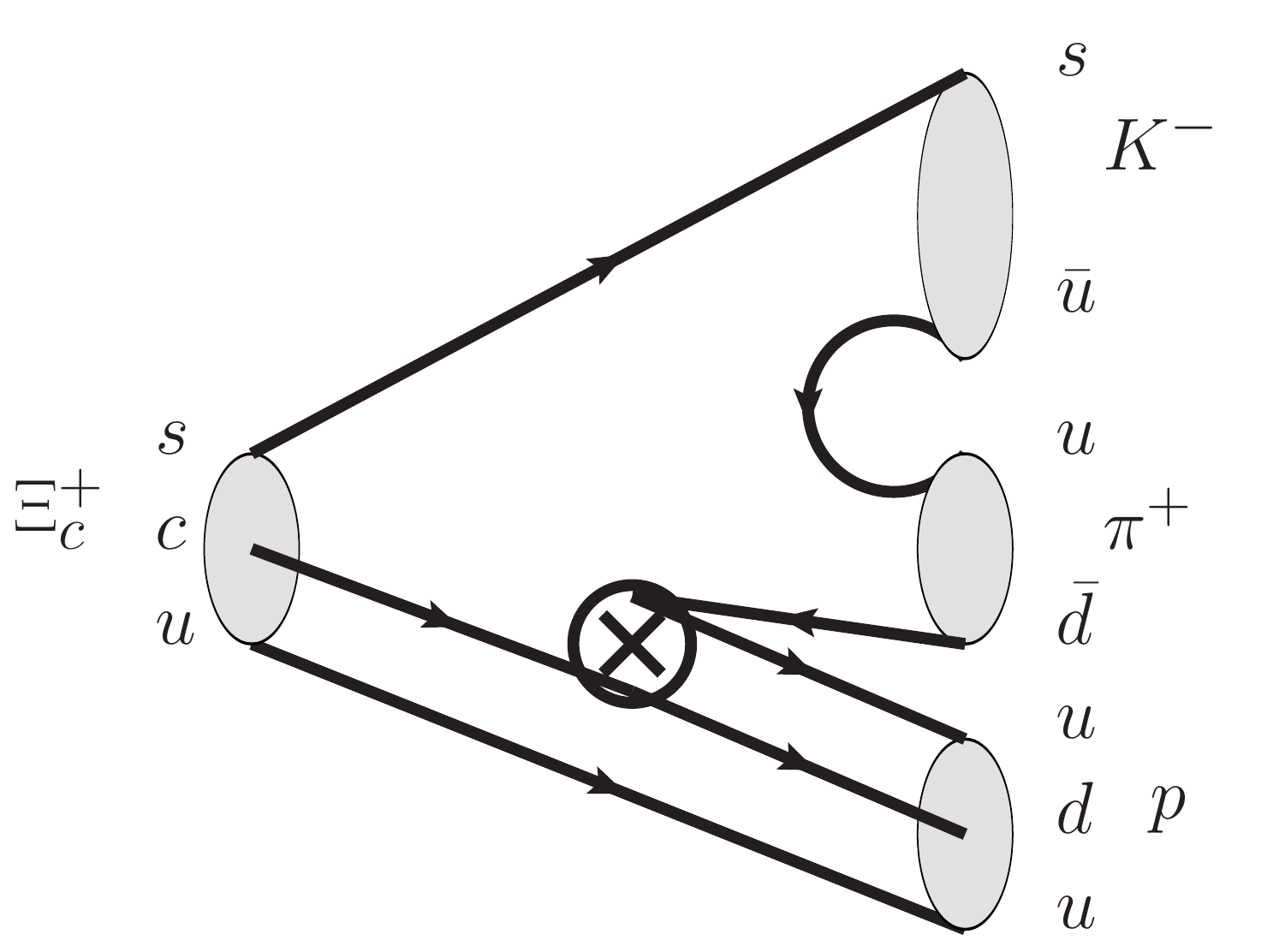}}
\subfigure[\,$E_1$]{\includegraphics[width=0.23\textwidth]{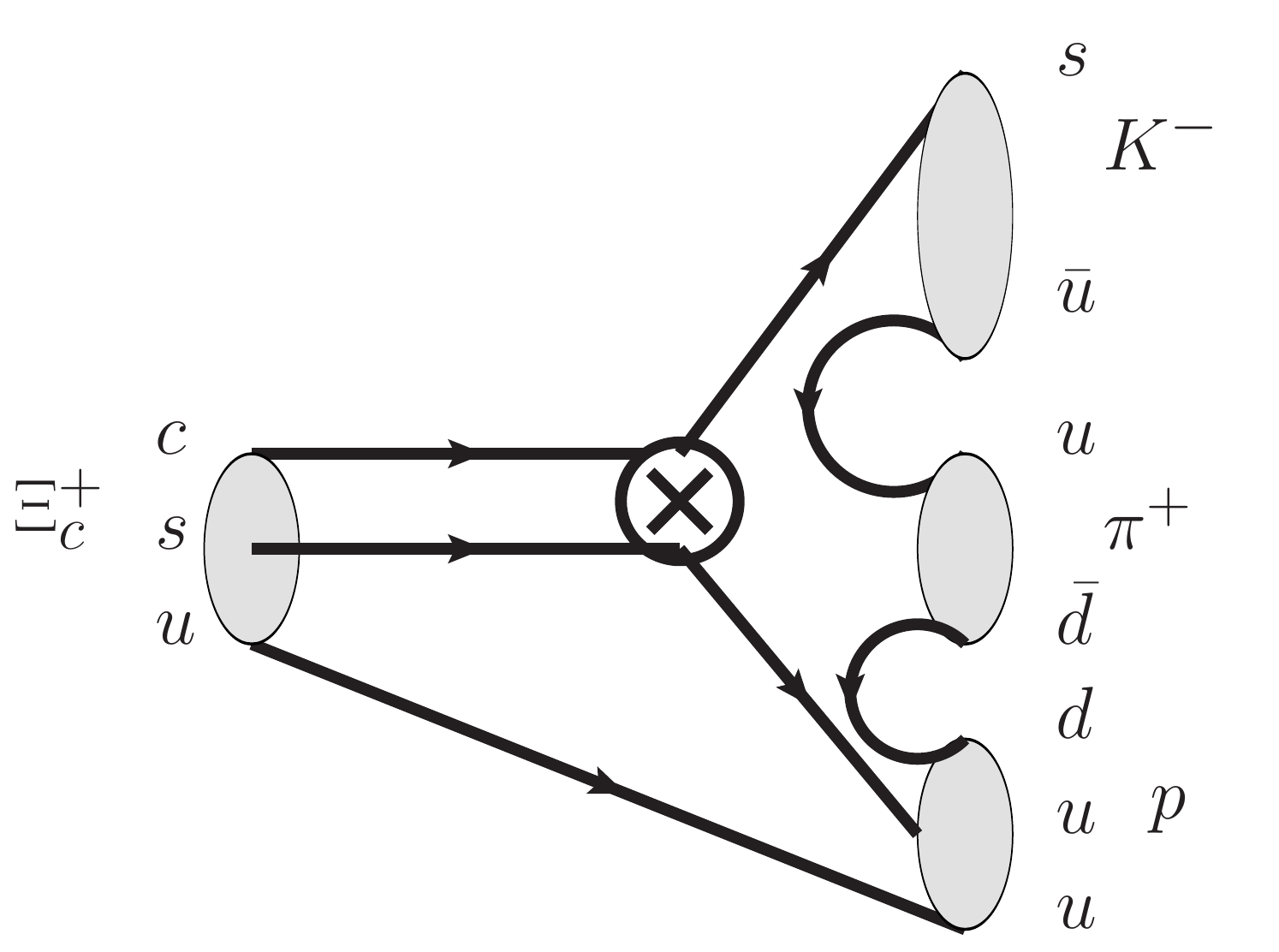}} 
\subfigure[\,$P_2$]{\includegraphics[width=0.23\textwidth]{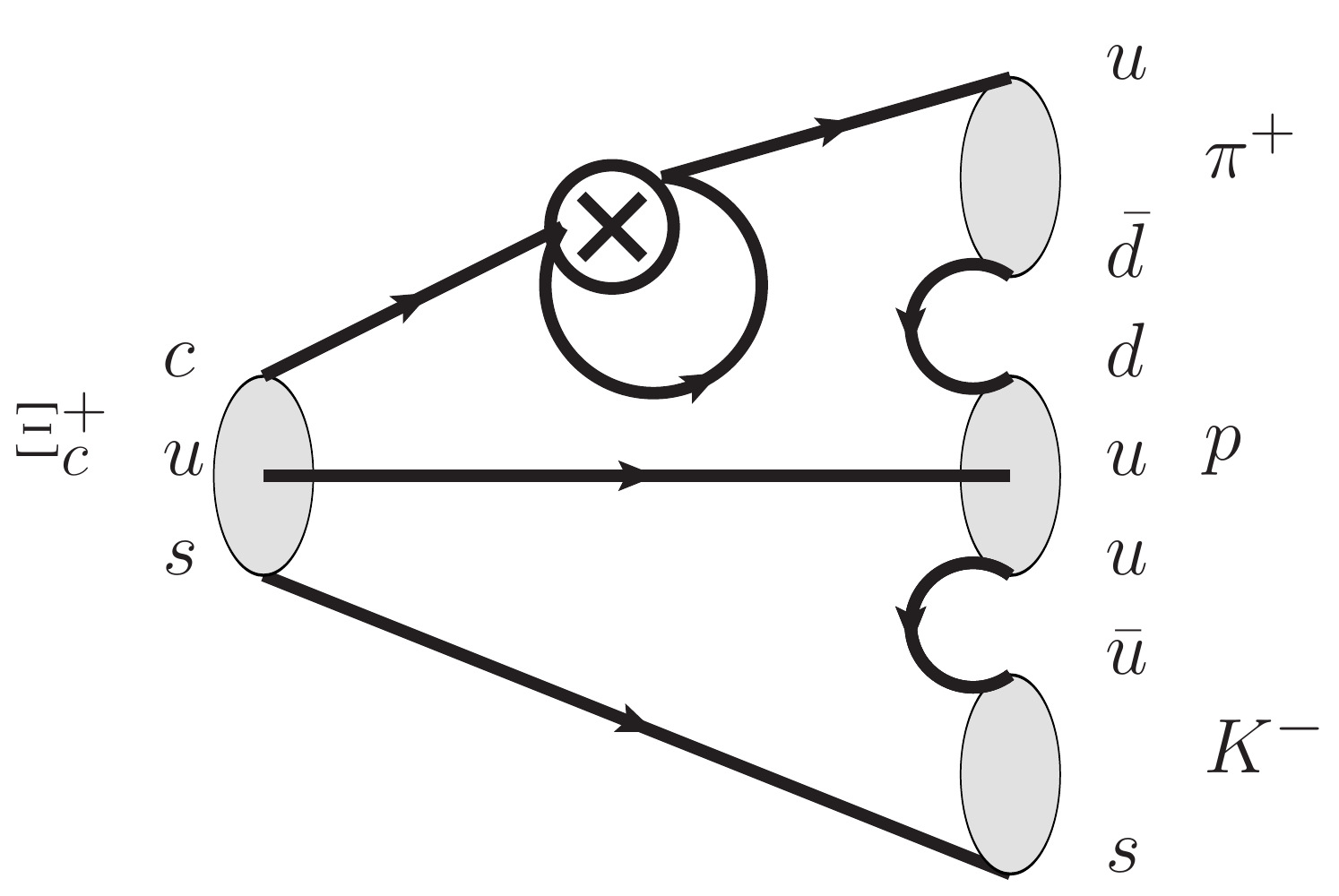}}
\caption{Diagrams in $\mathcal{A}(\Xi_c^+\rightarrow p K^- \pi^+) = (-\Sigma+\Delta) (-C_2) + (\Sigma+\Delta) (-E_1) + \Delta (-P_2) $. 
}
\label{fig:Xic-pKPi}
\end{center}
\end{figure}

\begin{figure}[h]
\begin{center}
\subfigure[\,$T$]{\includegraphics[width=0.23\textwidth]{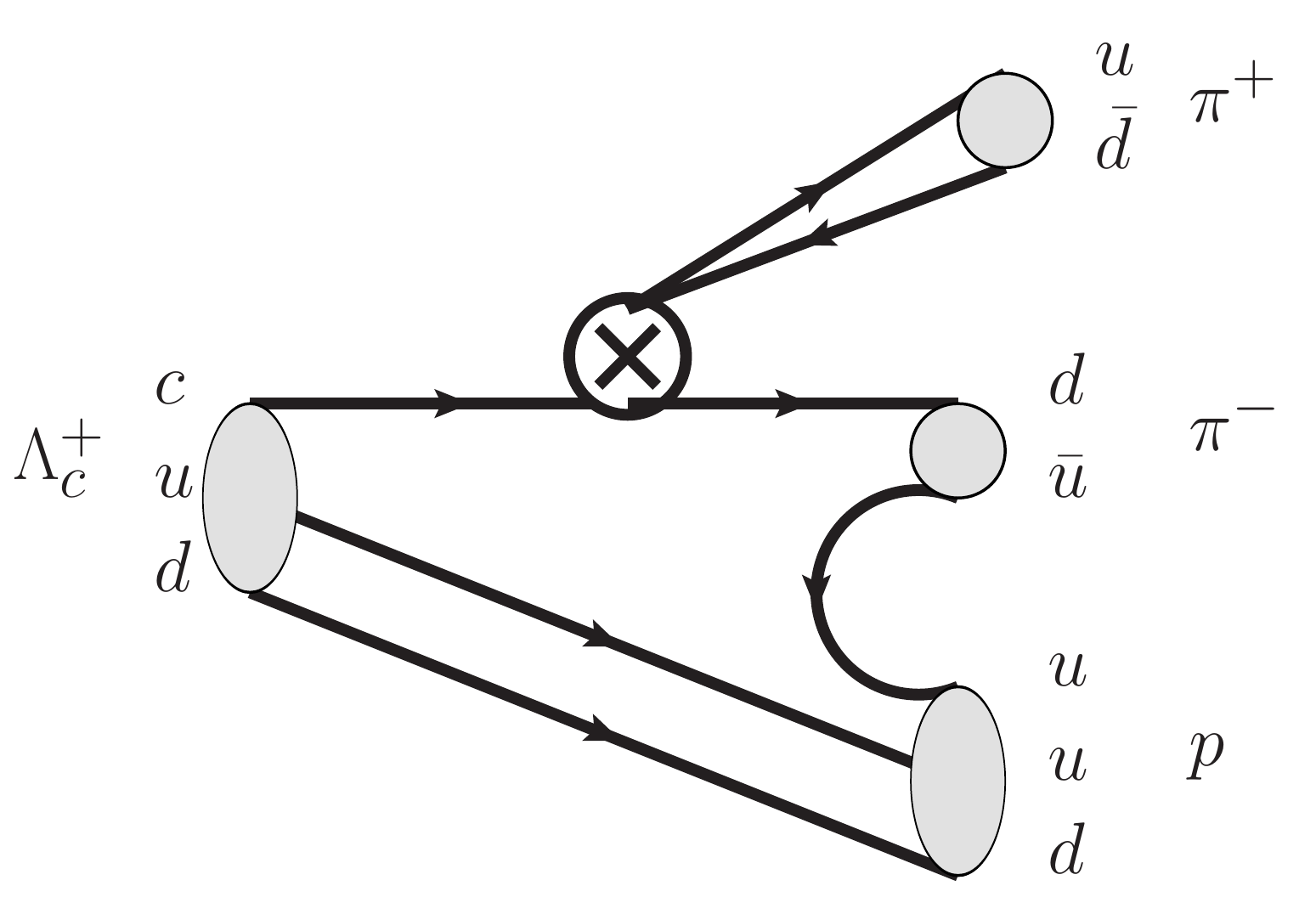}}
\subfigure[\,$C_1$]{\includegraphics[width=0.23\textwidth]{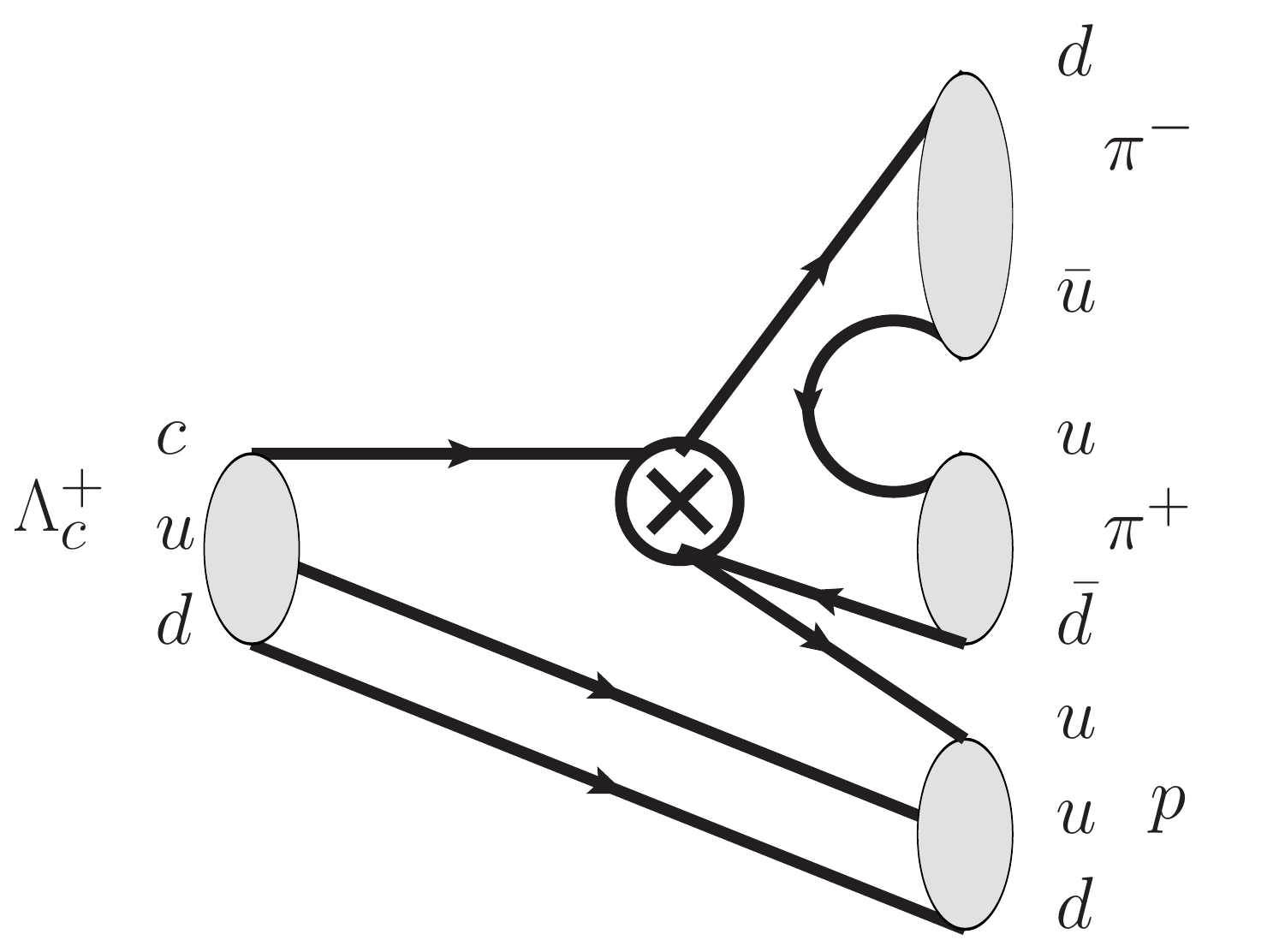}} 
\subfigure[\,$C_2$]{\includegraphics[width=0.23\textwidth]{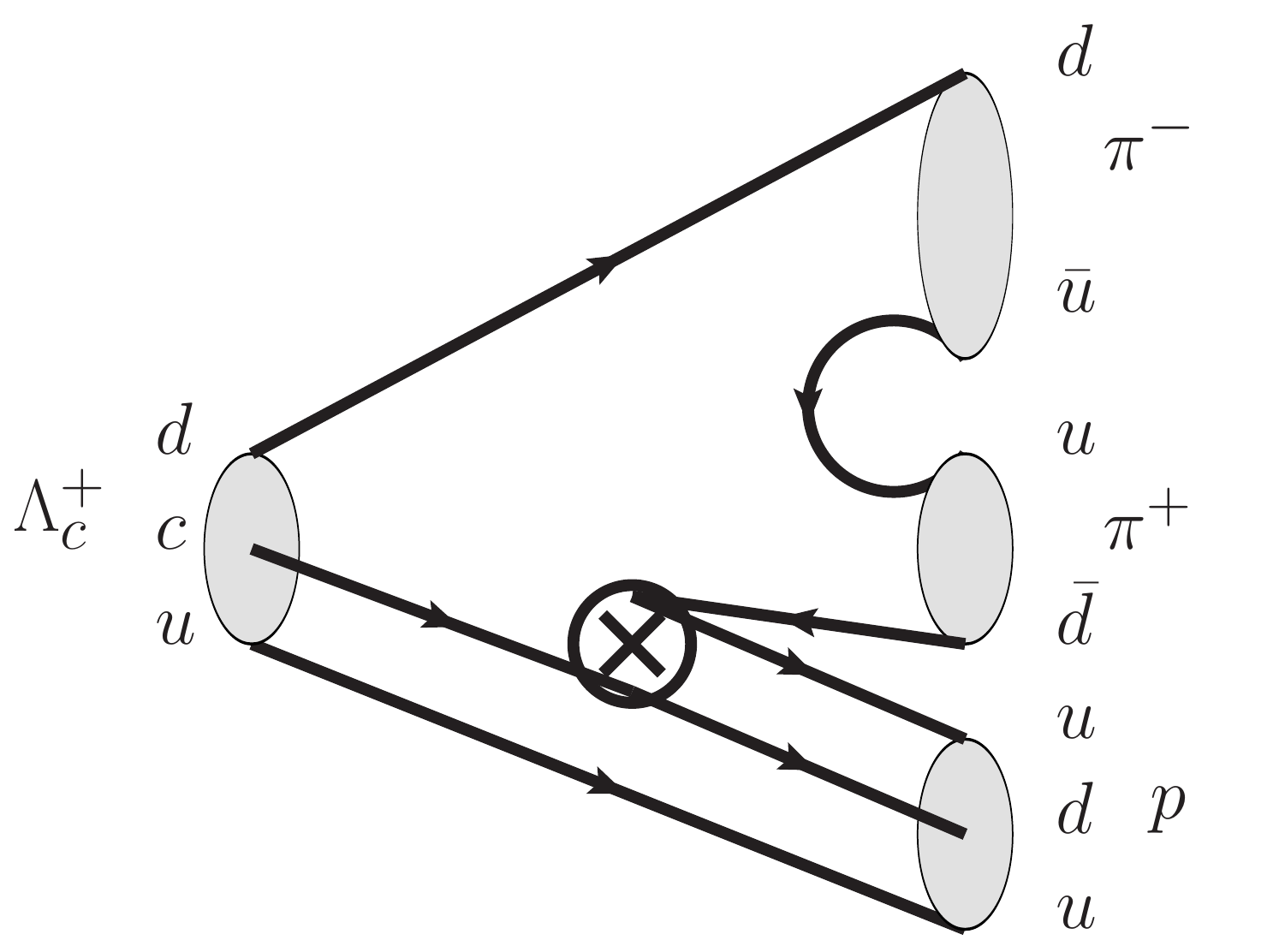}}
\subfigure[\,$E_1$]{\includegraphics[width=0.23\textwidth]{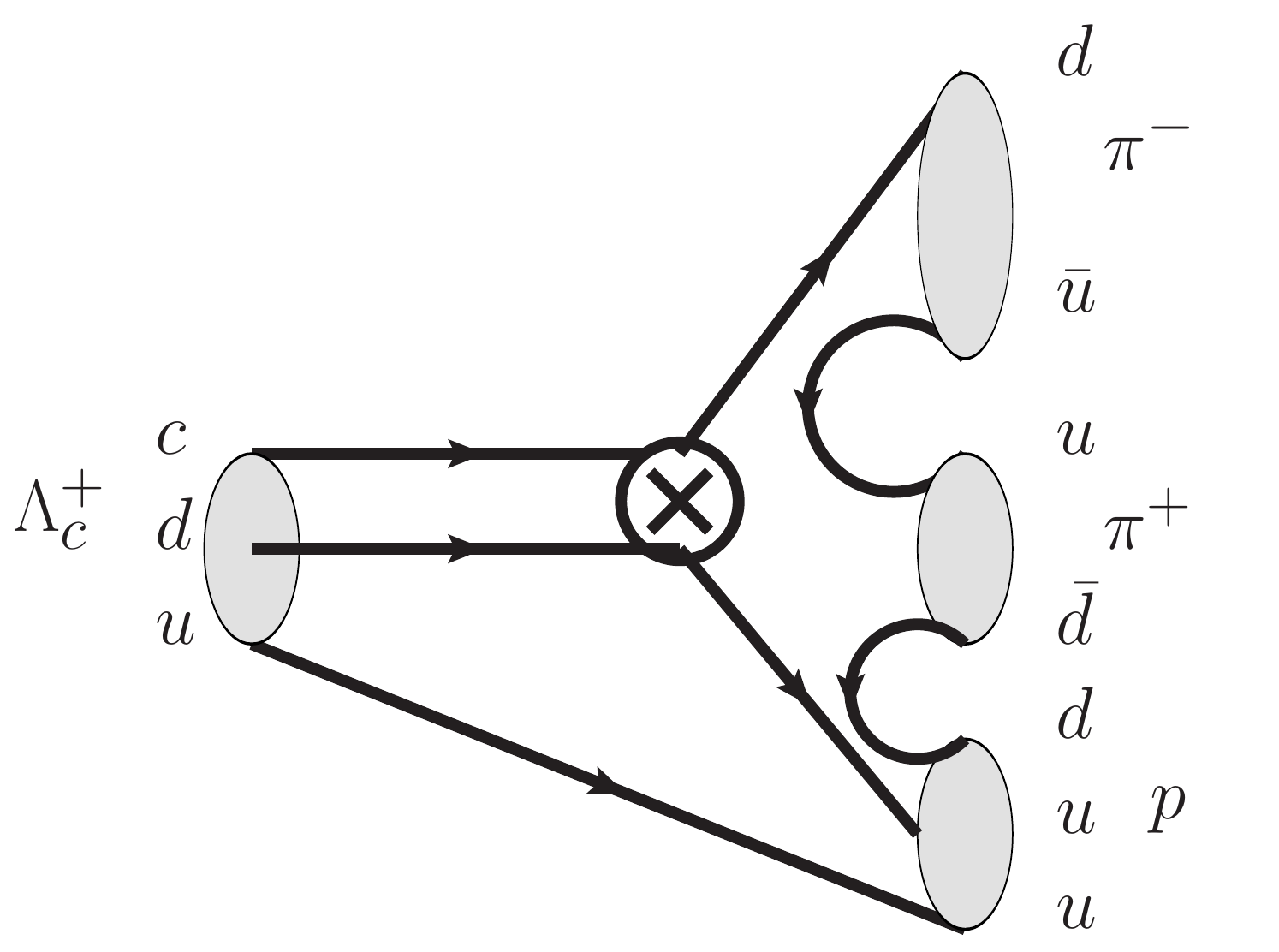}}
\subfigure[\,$E_2$]{\includegraphics[width=0.23\textwidth]{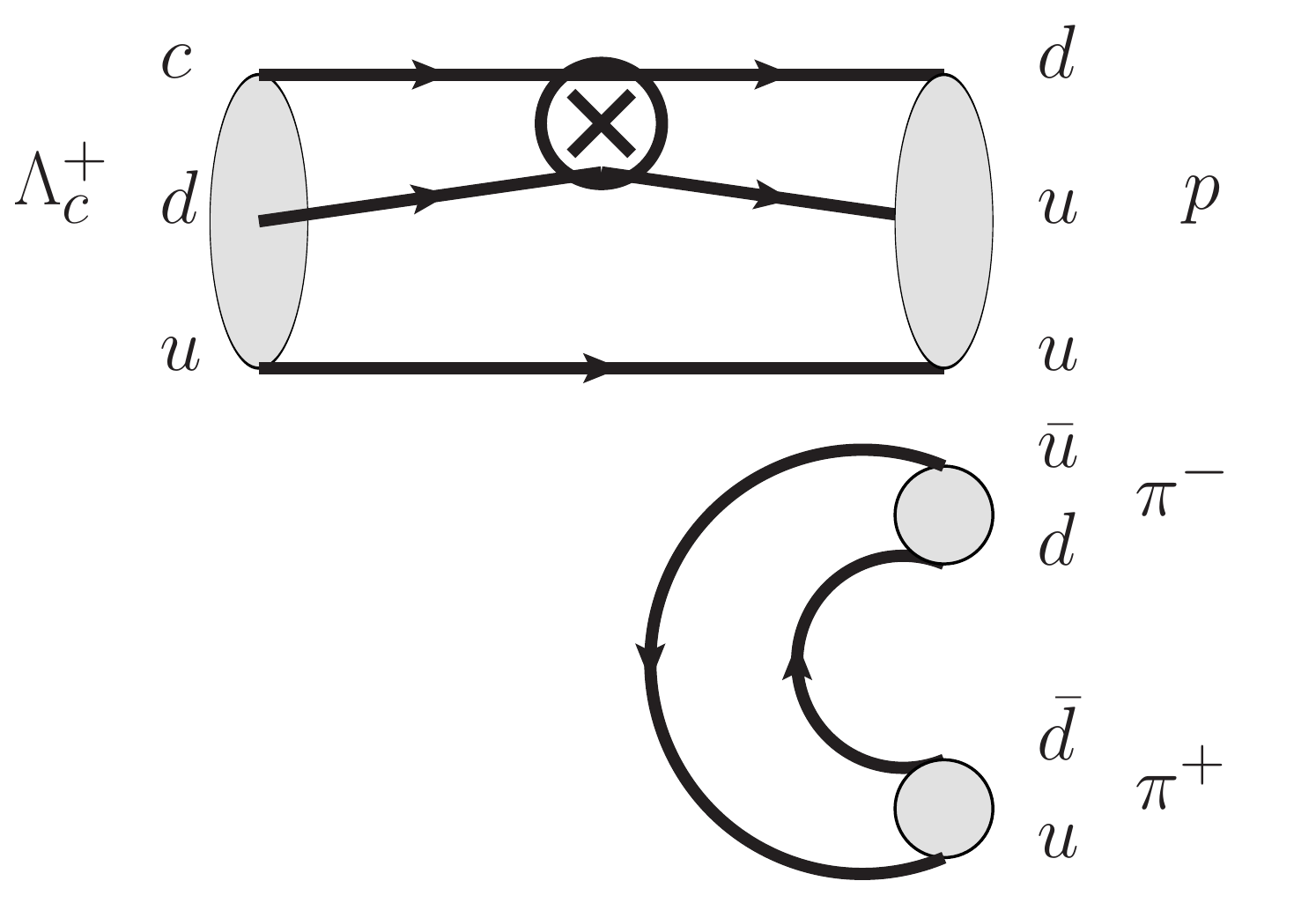}}
\subfigure[\,$P_1$]{\includegraphics[width=0.23\textwidth]{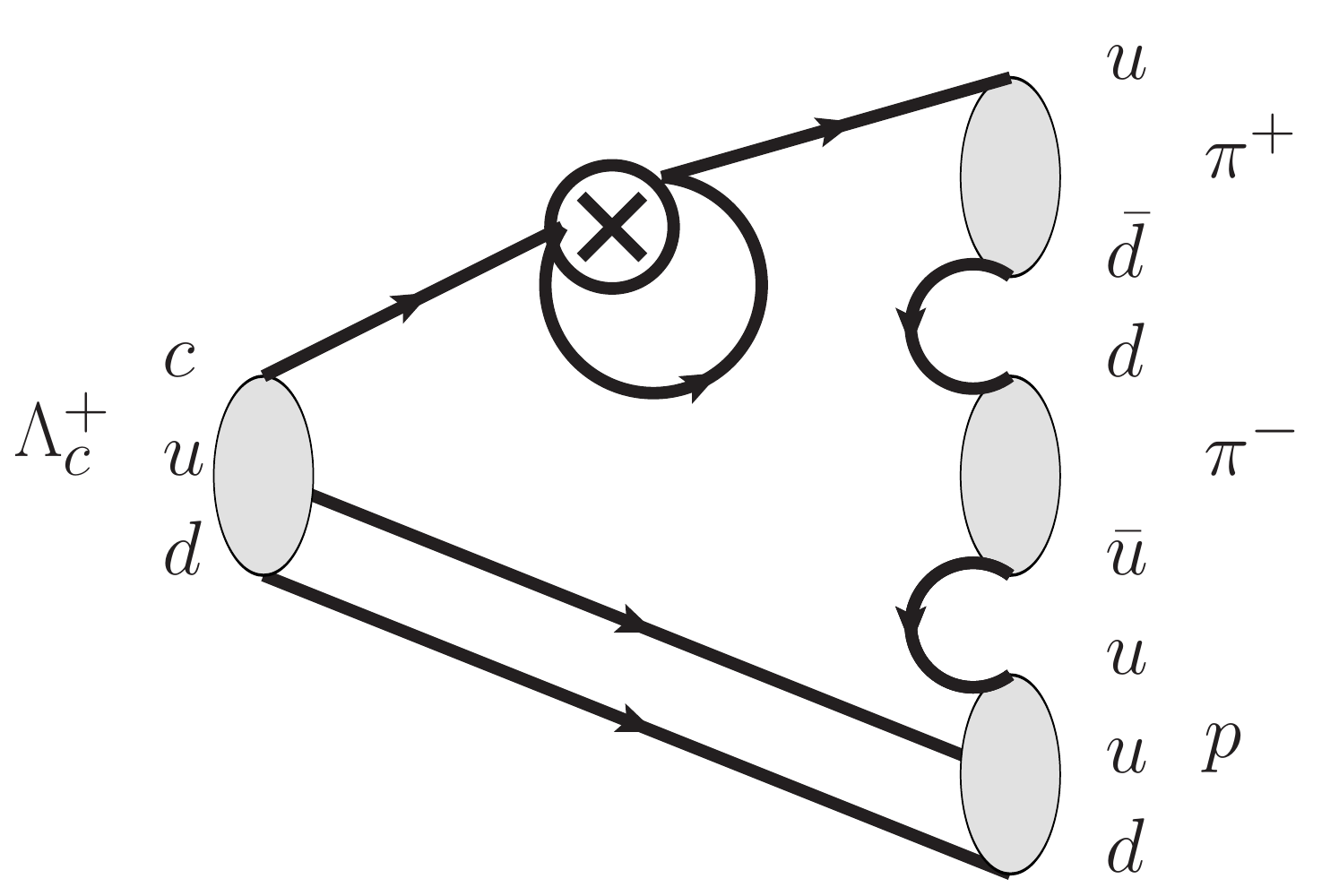}}
\subfigure[\,$P_2$]{\includegraphics[width=0.23\textwidth]{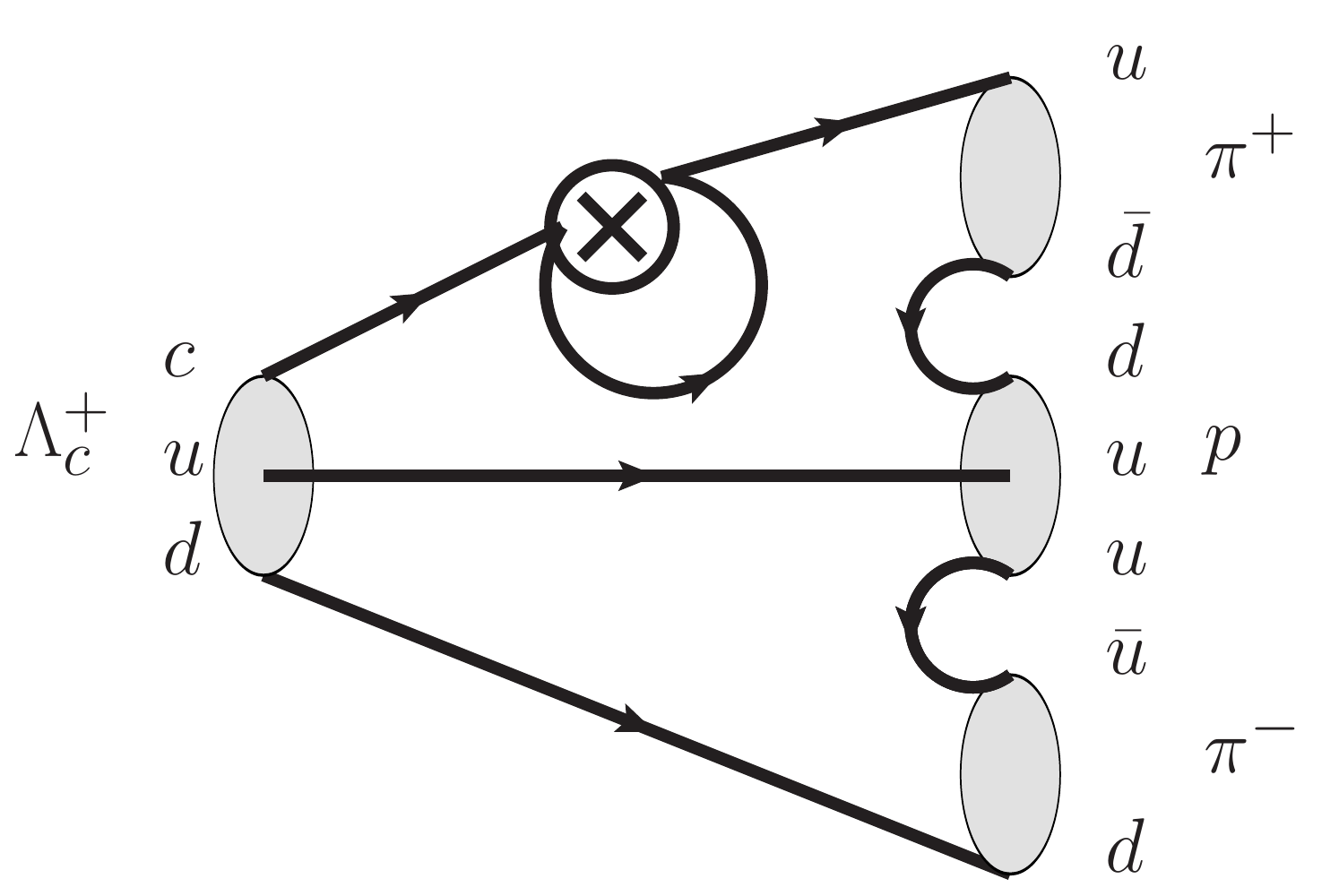}}
\caption{Diagrams in $\mathcal{A}(\Lambda_c^+\rightarrow p\pi^-\pi^+) =   
 	(-\Sigma+\Delta) (-T - C_1 - C_2 - E_1 - E_2 ) +\Delta (-P_1 - P_2)$.
}
\label{fig:Lambdac-pPiPi}
\end{center}
\end{figure}

\begin{figure}[h]
\begin{center}
\subfigure[\,$T$]{\includegraphics[width=0.23\textwidth]{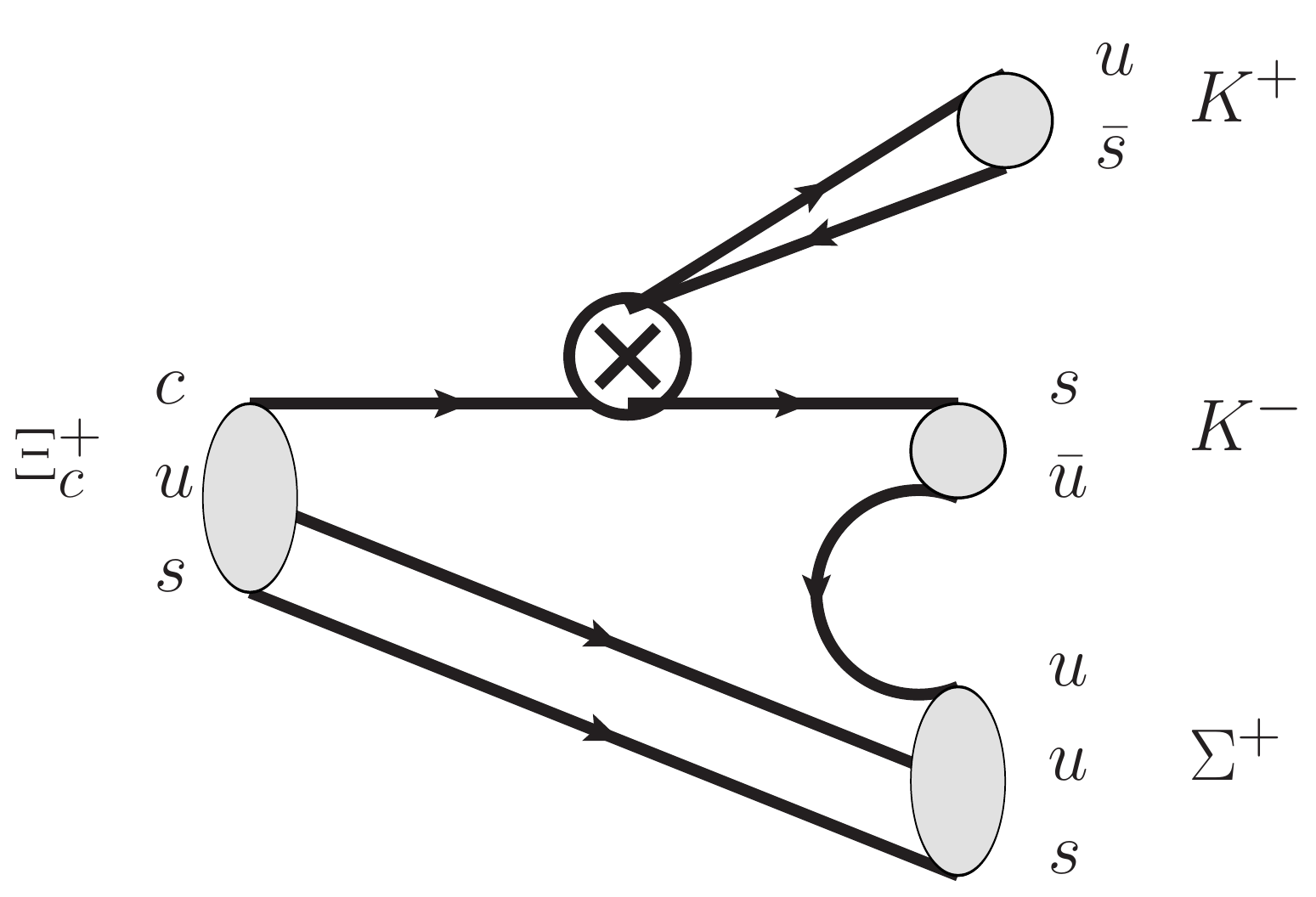}}
\subfigure[\,$C_1$]{\includegraphics[width=0.23\textwidth]{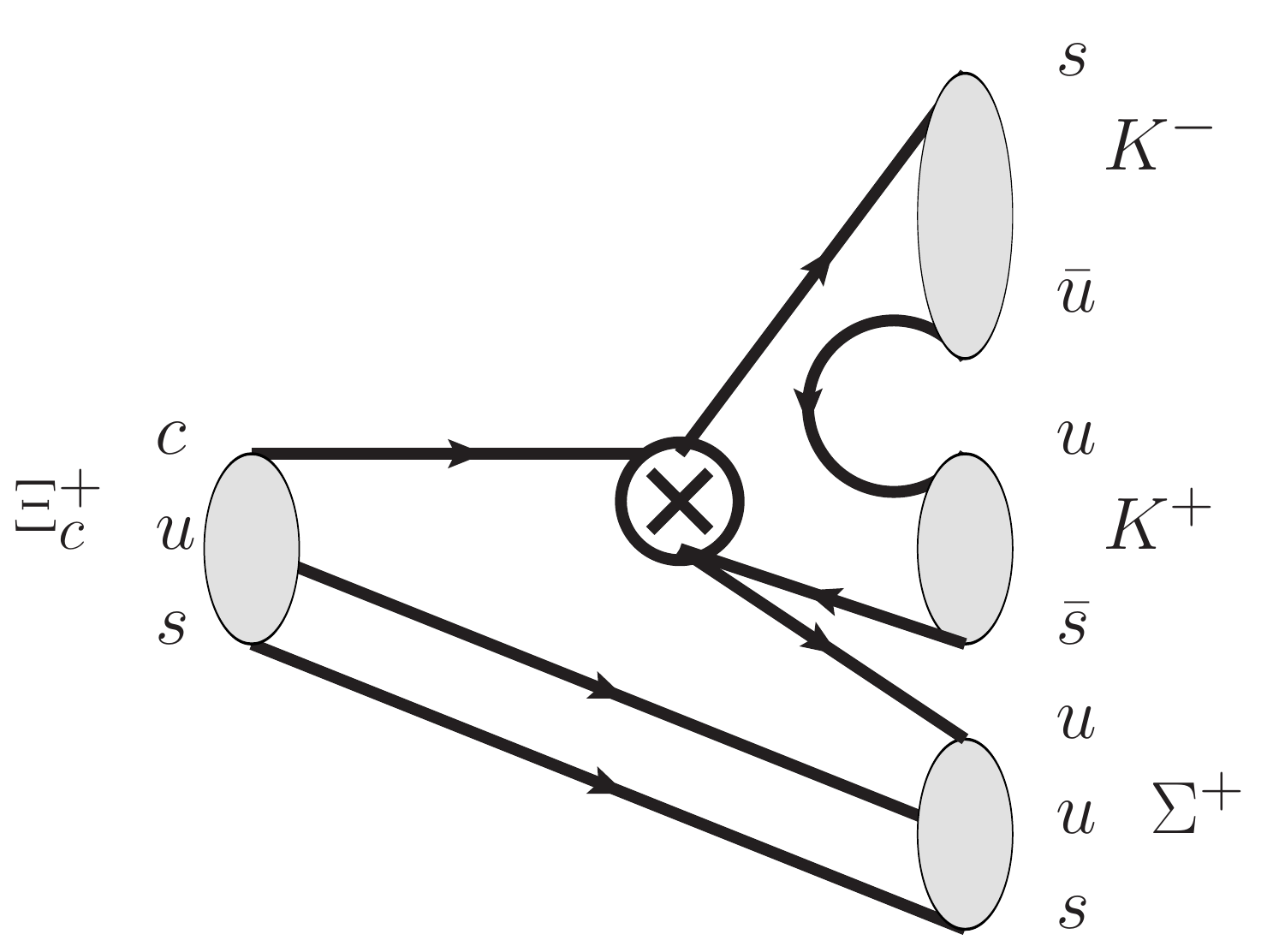}} 
\subfigure[\,$C_2$]{\includegraphics[width=0.23\textwidth]{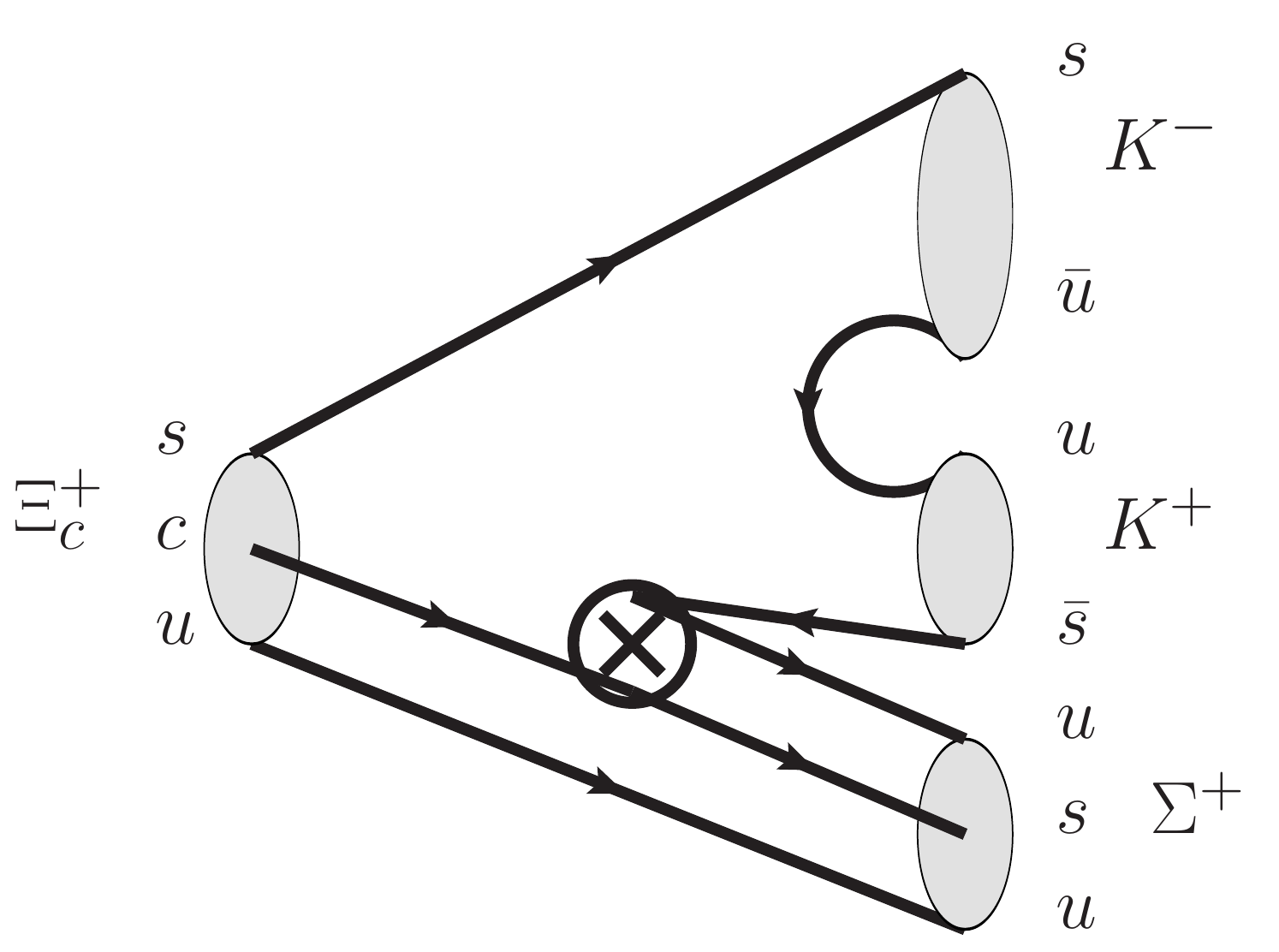}}
\subfigure[\,$E_1$]{\includegraphics[width=0.23\textwidth]{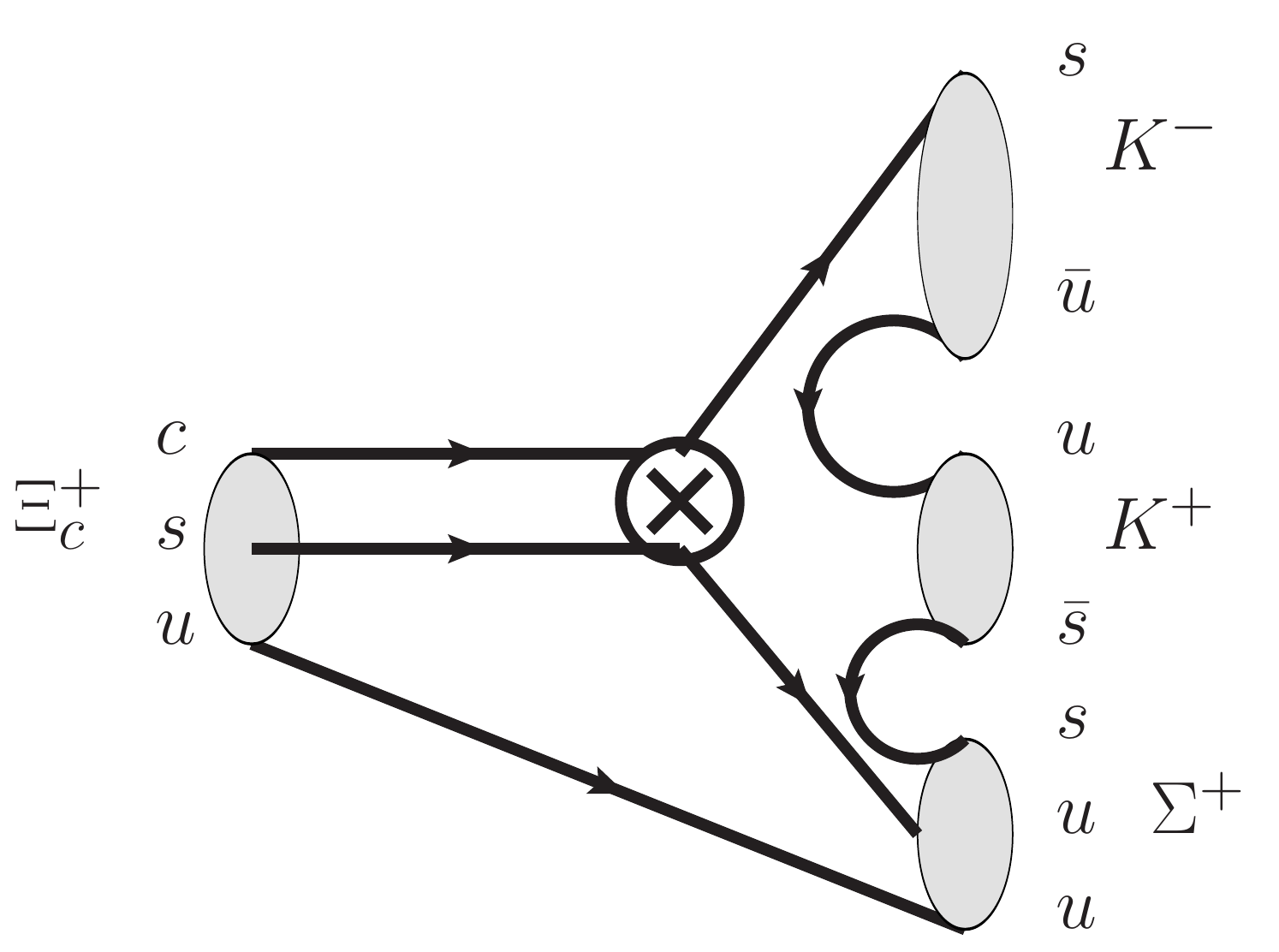}}
\subfigure[\,$E_2$]{\includegraphics[width=0.23\textwidth]{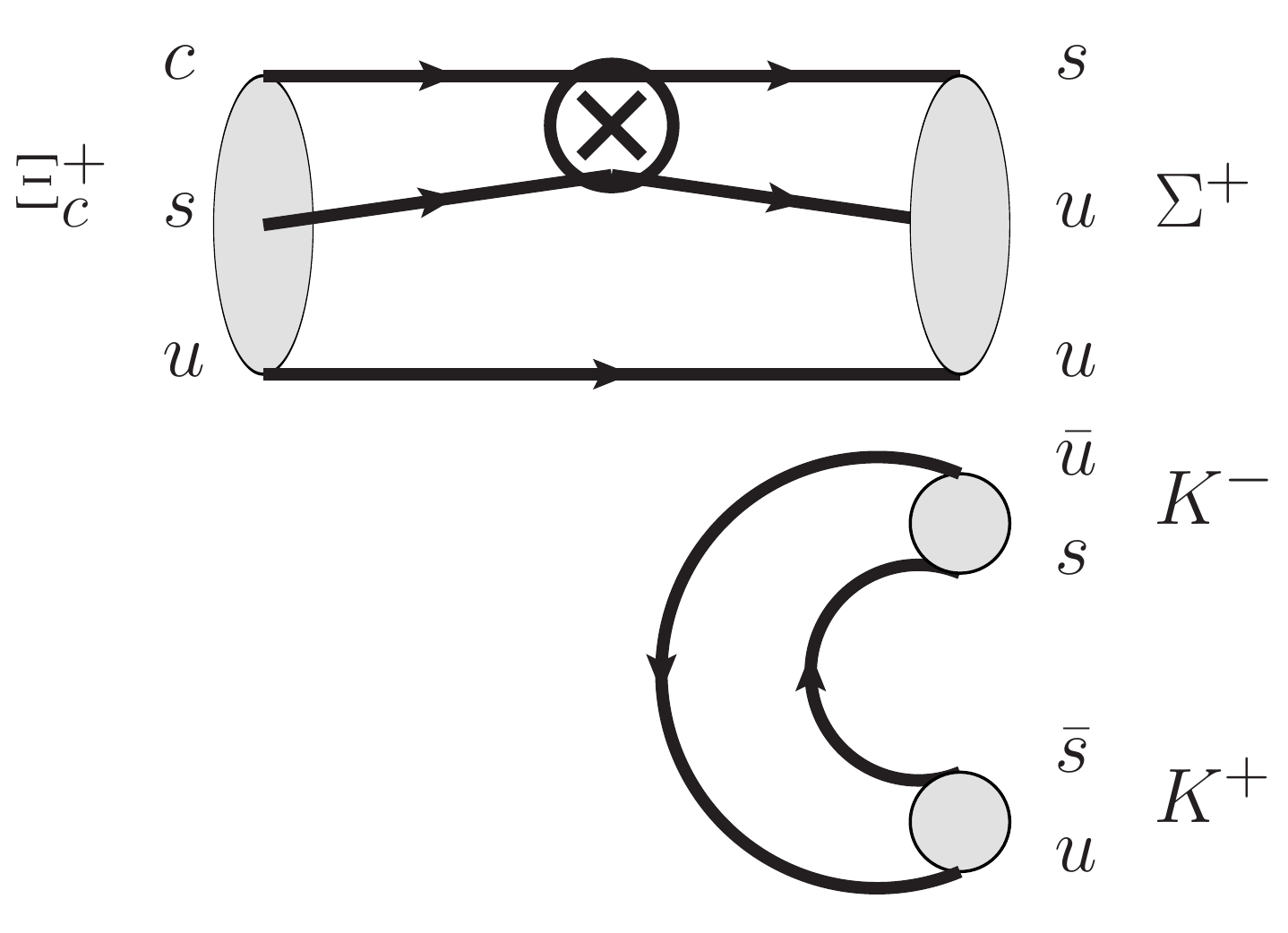}}
\subfigure[\,$P_1$]{\includegraphics[width=0.23\textwidth]{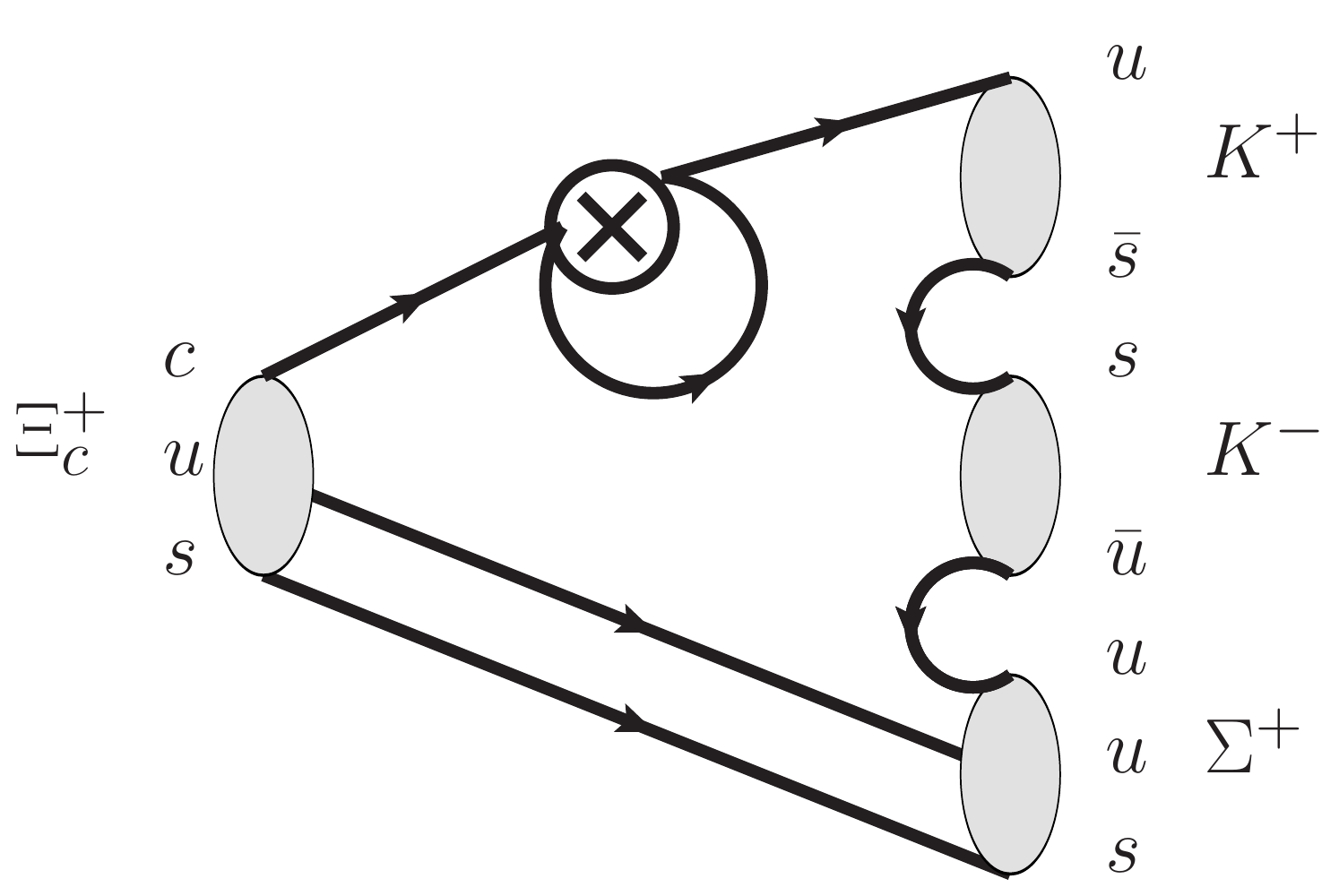}}
\subfigure[\,$P_2$]{\includegraphics[width=0.23\textwidth]{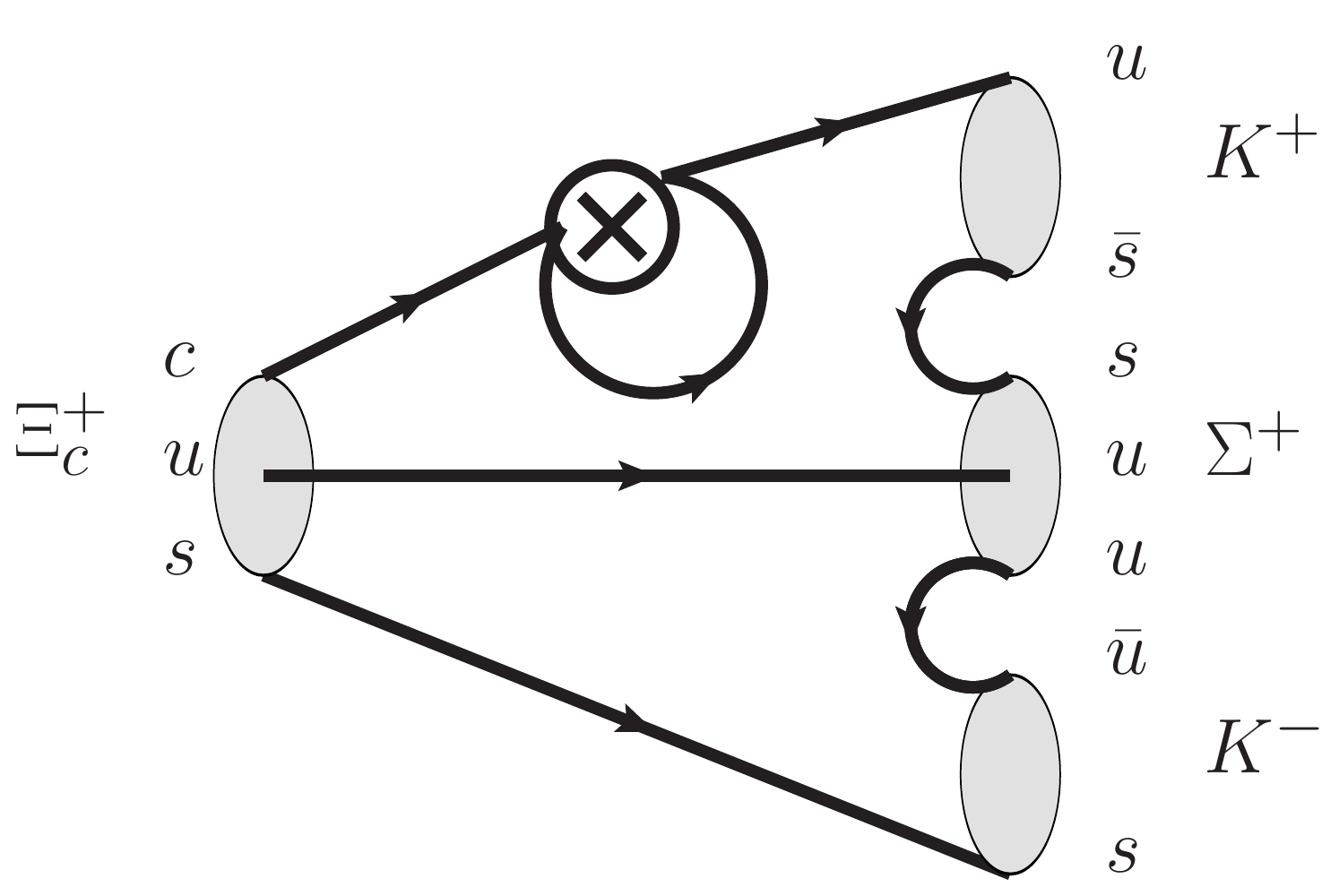}}
\caption{Diagrams in 
$\mathcal{A}(\Xi_c^+\rightarrow \Sigma^+K^-K^+) = (\Sigma+\Delta) (-T - C_1 - C_2 - E_1 - E_2) +\Delta (-P_1 - P_2)$.
}
\label{fig:Xic-SigmaKK}
\end{center}
\end{figure}

For the diagrammatic approach, the topological diagrams are shown in Figs.~\ref{fig:Lambdac-pKK}-\ref{fig:Xic-SigmaKK}.
The topological diagrams are all-order QCD diagrams which capture the flavor-flow only.
In each diagram we imply the sum over all possible combinations to connect the final state up quarks.
As we consider U-spin partners only here, these are the same for all decay channels.
Furthermore, in case of the penguin diagram the shown topology is defined as 
\begin{align}
P\equiv P_s+P_d-2P_b\,, 
\end{align}
where $P_q$ is the penguin diagram with the down-type quark $q$ running in the loop,
see Eq.~(\ref{eq:CKMstructure}) and Ref.~\cite{Muller:2015rna}.
Annihilation diagrams with antiquarks from the sea of the initial state do not play a role here. 
Our result for the diagrammatical decomposition is given in Table~\ref{tab:topo}, where  
we form combinations of the topologies which give linear independent contributions.
Note that the parametrizations in Tables~\ref{tab:group} and \ref{tab:topo} are equivalent, see also 
Refs.~\cite{Zeppenfeld:1980ex,Gronau:1994rj,Muller:2015lua} for the same observation for two-body meson decays.
Both of the shown parametrization matrices have rank five, and we have a one-to-one matching of the independent parameter combinations
of the two parametrizations on each other.
Explicitly, the mapping of the two parametrizations reads 
\begin{align}
\vect{
\bra{\frac{1}{2}}_0 1 \ket{\frac{1}{2}} \\ 
\bra{\frac{1}{2}}_1 1 \ket{\frac{1}{2}} \\
\bra{\frac{3}{2}}   1 \ket{\frac{1}{2}}   \\
\bra{\frac{1}{2}}_0 0 \ket{\frac{1}{2}} \\
\bra{\frac{1}{2}}_1 0 \ket{\frac{1}{2}} 
} = \left(
\begin{array}{ccccc}
 \sqrt{\frac{3}{2}} & -\sqrt{\frac{3}{2}} & \sqrt{\frac{3}{2}} & 0 & 0 \\
 \frac{1}{\sqrt{2}} & \frac{1}{\sqrt{2}} & \frac{3}{\sqrt{2}} & 0 & 0 \\
 \sqrt{2} & \sqrt{2} & 0 & 0 & 0 \\
 0 & 0 & 0 & \frac{1}{\sqrt{2}} & -\frac{1}{\sqrt{2}} \\
 0 & 0 & 0 & -\sqrt{\frac{3}{2}} & -\sqrt{\frac{3}{2}} \\
\end{array}
\right)
\vect{
 T + C_1 - E_2 \\
 C_2 + E_2 \\
 E_1 + E_2 \\
 T+C_1+E_2+P_1 \\
 C_2+E_1+P_2
 }\,. \label{eq:translation}
\end{align}
Herein, the first three reduced matrix elements correspond to the CKM-leading part, and the last two to the CKM-subleading part, 
which is why the translation matrix is block diagonal.
As the coefficient submatrix of the CKM-leading part has matrix rank three, there are three U-spin sum rules for the $A_{\Sigma}$
part, which one can read off directly as  
\begin{align}
A_{\Sigma}(\Lambda_c^+\rightarrow pK^-K^+) &= -A_{\Sigma}(\Xi_c^+\rightarrow \Sigma^+ \pi^- \pi^+)\,,    \label{amp-sum-rule-1}\\
A_{\Sigma}(\Lambda_c^+\rightarrow \Sigma^+\pi^-K^+) &= -A_{\Sigma}(\Xi_c^+\rightarrow p K^- \pi^+) \,,   \label{amp-sum-rule-2} \\
A_{\Sigma}(\Lambda_c^+\rightarrow p \pi^- \pi^+) &= -A_{\Sigma}(\Xi_c^+\rightarrow \Sigma^+ K^- K^+)\,.  \label{amp-sum-rule-3}
\end{align}
The sum rules Eqs.~(\ref{amp-sum-rule-1})--(\ref{amp-sum-rule-3}) agree with the ones that one can read off Table~XI in 
Ref.~\cite{Savage:1989qr}.
The CKM-subleading coefficient submatrix has rank two, so there are four corresponding U-spin sum rules.
Three of them correspond to the ones for the CKM-leading part, namely
\begin{align}
A_{\Delta}(\Lambda_c^+\rightarrow pK^-K^+)          &= A_{\Delta}(\Xi_c^+\rightarrow \Sigma^+ \pi^- \pi^+)\,, \label{amp-sum-rule-4}\\
A_{\Delta}(\Lambda_c^+\rightarrow \Sigma^+\pi^-K^+) &= A_{\Delta}(\Xi_c^+\rightarrow p K^- \pi^+) \,, 	      \label{amp-sum-rule-5} \\
A_{\Delta}(\Lambda_c^+\rightarrow p \pi^- \pi^+)    &= A_{\Delta}(\Xi_c^+\rightarrow \Sigma^+ K^- K^+)\,.     \label{amp-sum-rule-6}
\end{align}
The additional one is given as 
\begin{align}
A_{\Delta}(\Lambda_c^+\rightarrow pK^-K^+) + A_{\Delta}(\Lambda_c^+\rightarrow \Sigma^+\pi^-K^+) = 
A_{\Delta}(\Lambda_c^+\rightarrow p \pi^- \pi^+)\,.
\end{align}
Finally, the full U-spin limit coefficient matrix has rank five, therefore there is one sum rule for the full amplitudes
\begin{align}
&\mathcal{A}(\Lambda_c^+\rightarrow p K^-K^+)
+\mathcal{A}(\Xi_c^+\rightarrow \Sigma^+ \pi^-\pi^+)
+\mathcal{A}(\Lambda_c^+\rightarrow \Sigma^+\pi^-K^+)+\nn\\
&\mathcal{A}(\Xi_c^+\rightarrow p K^- \pi^+) 
-\mathcal{A}(\Lambda_c^+\rightarrow p \pi^-\pi^+)
-\mathcal{A}(\Xi_c^+\rightarrow \Sigma^+ K^-K^+) = 0\,.
\end{align} 

\section{CP Asymmetry Sum Rules \label{sec:sumrules}} 

We start our discussion in the U-spin limit (later we consider also U-spin breaking). 
Furthermore, as~\cite{Tanabashi:2018oca}
\begin{align}
\mathrm{Im}\left(-2\Delta/\Sigma\right)\approx -6\cdot10^{-4}, 
\end{align}
disregarding powers of $\mathcal{O}\left(\Delta^2/\Sigma^2\right)$ is an excellent approximation.
Within this approximation, the CP asymmetry at a certain point in the Dalitz plot can be 
written as (see, \emph{e.g.},~Refs.~\cite{Golden:1989qx, Pirtskhalava:2011va, Nierste:2017cua}) 
\begin{align}
a_{CP} &\equiv 
\frac{\vert \mathcal{A}\vert^2 - \vert \bar{\mathcal{A}}\vert^2 }{ 
      \vert \mathcal{A}\vert^2 + \vert \bar{\mathcal{A}}\vert^2 }
= \mathrm{Im}\left(\frac{-2\Delta}{\Sigma}\right) \mathrm{Im}\left(\frac{A_{\Delta}}{A_{\Sigma}}\right)\,. \label{eq:CPasym}
\end{align}
Inserting the amplitude sum rules Eqs.~(\ref{amp-sum-rule-1})--(\ref{amp-sum-rule-6}) into Eq.~(\ref{eq:CPasym}) we obtain the pointwise CP asymmetry sum rules
\begin{align}
a_{CP}(\Lambda_c^+\rightarrow p K^- K^+) + 
a_{CP}(\Xi_c^+\rightarrow \Sigma^+ \pi^- \pi^+) &= 0\,, \label{eq:sum-rule-1}\\
a_{CP}(\Lambda_c^+\rightarrow \Sigma^+\pi^-K^+ ) + 
a_{CP}(\Xi_c^+\rightarrow p K^- \pi^+) &= 0\,, \label{eq:sum-rule-2}\\
a_{CP}(\Lambda_c^+\rightarrow p \pi^- \pi^+) + 
a_{CP}(\Xi_c^+\rightarrow \Sigma^+ K^- K^+) &= 0\,. \label{eq:sum-rule-3}
\end{align}
 
Next, we move to the discussion of the phase space integrated CP asymmetry. 
In the U-spin limit and to linear order in $\Delta/\Sigma$ it is given as 
\begin{align}
A_{CP} &\equiv 
\frac{\int \vert \mathcal{A}\vert^2 dp - \int \vert \bar{\mathcal{A}}\vert^2dp }{ 
      \int \vert \mathcal{A}\vert^2 dp + \int \vert \bar{\mathcal{A}}\vert^2 dp }  
= \mathrm{Im}\left(\frac{-2\Delta}{\Sigma}\right) I_p 
   \,, \label{eq:phase-space-integrated-ACP} 
\end{align}
with
\begin{align}
I_p &= \frac{\int \mathrm{Im}\left(A_{\Sigma}^*A_{\Delta}\right) dp}{ 
	\int \vert A_{\Sigma}\vert^2 dp
	}\,.
\end{align}
Here, the $dp$-integration denotes the integration over all phase space variables. 

In case of two-body charm meson decays to pseudoscalars, Eq.~(\ref{eq:phase-space-integrated-ACP}) gives a trivial integral 
and we have $A_{CP} = a_{CP}$ as it must be.
Note that for $D^0$ decays the CP asymmetries have additional
contributions from indirect CP violation due to charm mixing. 
This additional complication is not present for baryon decay. 

In order to promote a sum rule which is valid for pointwise CP asymmetries, $a_{CP}$,
to a sum rule between 
CP asymmetries of integrated rates, $A_{CP}$, it is necessary that $\vert I_p\vert$ agrees for the involved CP asymmetries. 
From Eqs.~(\ref{amp-sum-rule-1})--(\ref{amp-sum-rule-6}) it is clear that this criterion is fulfilled by all 
three pairs of decays in Eqs.~(\ref{eq:sum-rule-1})--(\ref{eq:sum-rule-3}).
Thus, the pointwise sum rules can be promoted to ones for CP asymmetries of the integrated rates 
\begin{align}
A_{CP}(\Lambda_c^+\rightarrow p K^- K^+) + 
A_{CP}(\Xi_c^+\rightarrow \Sigma^+ \pi^- \pi^+) &= 0\,, \label{eq:sum-rule-4}\\
A_{CP}(\Lambda_c^+\rightarrow \Sigma^+\pi^-K^+ ) + 
A_{CP}(\Xi_c^+\rightarrow p K^- \pi^+) &= 0\,, \label{eq:sum-rule-5}\\
A_{CP}(\Lambda_c^+\rightarrow p \pi^- \pi^+) + 
A_{CP}(\Xi_c^+\rightarrow \Sigma^+ K^- K^+) &= 0\,. \label{eq:sum-rule-6}
\end{align}
Moreover, from Tables~\ref{tab:group} and \ref{tab:topo} it is clear that no such sum rule 
connects $A_{CP}(\Lambda_c^+\rightarrow p K^- K^+)$ and $A_{CP}(\Lambda_c^+\rightarrow p \pi^- \pi^+)$.
Additionally, as we discuss in Appendix~\ref{sec:uspin-breaking}, there are not even pointwise CP asymmetry sum rules at first order U-spin 
breaking. This means that Eqs.~(\ref{eq:sum-rule-1})--(\ref{eq:sum-rule-3}) and Eqs.~(\ref{eq:sum-rule-4})--(\ref{eq:sum-rule-6})
are expected to get corrections of $\mathcal{O}(30\%)$~\cite{Brod:2012ud,Hiller:2012xm,Muller:2015rna}.

\section{Conclusions \label{sec:conclusions}}

We construct U-spin CP asymmetry sum rules between SCS three-body charmed baryon decays,
which we give in Eqs.~(\ref{eq:sum-rule-1})--(\ref{eq:sum-rule-3}) and Eqs.~(\ref{eq:sum-rule-4})--(\ref{eq:sum-rule-6}).
The sum rules are valid both pointwise at any point in the Dalitz plot and for the integrated CP asymmetries.
There are no U-spin CP asymmetry sum rules besides the trivial ones due to the interchange of all $d$ and $s$ quarks.
Furthermore, there is no U-spin CP asymmetry sum rule which is valid beyond the U-spin limit. 
Also, there is no U-spin sum rule connecting $A_{CP}(\Lambda_c^+\rightarrow p K^-K^+)$ and 
$A_{CP}(\Lambda_c^+\rightarrow p \pi^-\pi^+)$ whose difference recently has been measured by LHCb~\cite{Aaij:2017xva}. 
The dynamic reason for the latter is that the presence of the spectator quark and the additional combinatorial possibilities 
due to the three-body decay lead eventually to more possible topological 
combinations for $\Lambda_c^+\rightarrow p \pi^-\pi^+$ than for the $\Lambda_c^+\rightarrow p K^-K^+$ in both 
the CKM-leading and the CKM-subleading parts of the amplitudes. 
These additional contributions remain in the sum of the two CP asymmetries and do not cancel out. 

There are more opportunities for studying U-spin sum rules and their breaking in three-body charm decays by including also 
the branching ratios of Cabibbo-favored and doubly Cabibbo-suppressed decays into the discussion, which we leave for future work.

\begin{acknowledgments}
We thank Alan Schwartz for asking the question which led to this work. 
The work of YG is supported in part by the NSF grant PHY1316222.
SS is supported by a DFG Forschungs\-stipendium under contract no. SCHA 2125/1-1.	
\end{acknowledgments}

\appendix

\section{U-spin breaking \label{sec:uspin-breaking}}

The U-spin breaking from the difference of $d$ an $s$ quark masses gives rise to a triplet spurion operator.
For implications for meson decays see,~\emph{e.g.},~Refs.~\cite{Brod:2012ud,Feldmann:2012js,Jung:2009pb,Fleischer:2016ofb}. 
In order to include these corrections within perturbation theory we perform the tensor products with the unperturbed Hamiltonian. 
We have
\begin{align}
(1,0) \otimes (1,0) = \sqrt{\frac{2}{3}} (2,0) - \sqrt{\frac{1}{3}} (0,0)\,. \label{eq:breakingoperator}
\end{align} 
Note that there is no triplet present on the right hand side in Eq.~(\ref{eq:breakingoperator}) as $\Delta U_3=0$ for both $\Delta U=1$ operators
on the left hand side and the $(1,0)$ in the corresponding product comes with a vanishing Clebsch-Gordan coefficient. 
Our result for the parametrization of the CKM-leading U-spin breaking contribution $\mathcal{A}_X$ to the decay amplitudes is given 
in Table~\ref{tab:groupbreaking}. 
Combining this result with the CKM-leading part of the parametrization given in Table~\ref{tab:group} we obtain a matrix with rank six. 
That means there are no U-spin sum rules valid at this order between the SCS decays---neither for the full amplitudes nor 
for the CKM-leading part only. Furthermore, at this order there are not even pointwise CP asymmetry sum rules, not to mention ones for 
CP asymmetries of integrated rates.

\newpage

\begin{table}[h]
\begin{center}
\begin{tabular}{c|c|c}
\hline \hline
Decay ampl. {$\mathcal{A}$} & $c\rightarrow s$  & $c\rightarrow d$ \\\hline\hline
$\mathcal{A}(\Lambda_c^+\rightarrow p K^-K^+)$   	& \checkmark & \checkmark \\
$\mathcal{A}(\Xi_c^+\rightarrow \Sigma^+ \pi^-\pi^+)$   & \checkmark & \checkmark \\
$\mathcal{A}(\Lambda_c^+\rightarrow \Sigma^+\pi^-K^+)$  & $\times$  & \checkmark \\
$\mathcal{A}(\Xi_c^+\rightarrow p K^-\pi^+)$  		& \checkmark & $\times$ \\   
$\mathcal{A}(\Lambda_c^+\rightarrow p \pi^-\pi^+)$   	& $\times$  & \checkmark \\
$\mathcal{A}(\Xi_c^+\rightarrow \Sigma^+ K^-K^+)$  	& \checkmark & $\times$ \\\hline\hline
\end{tabular}
\caption{SCS decays connected to $\Lambda_c\rightarrow p\pi^+\pi^-$ by U-spin and their underlying quark level transitions in the non-penguin diagrams. \label{tab:quarklevel}}
\end{center}
\end{table}

\begin{table}[h]
\begin{center}
\begin{tabular}{c|c|c|c|c|c}
\hline \hline
Decay ampl. {$\mathcal{A}$} & 
$\Sigma \bra{\frac{1}{2}}_0  1 \ket{\frac{1}{2}}$ &  
$\Sigma \bra{\frac{1}{2}}_1  1 \ket{\frac{1}{2}}$ &  
$\Sigma \bra{\frac{3}{2}}    1 \ket{\frac{1}{2}}$ &  
$\Delta \bra{\frac{1}{2}}_0  0 \ket{\frac{1}{2}}$ &  
$\Delta \bra{\frac{1}{2}}_1  0 \ket{\frac{1}{2}}$  \\[6pt]\hline\hline
$\mathcal{A}(\Lambda_c^+\rightarrow p K^-K^+)$   & $-\frac{1}{\sqrt{6}}$  & $\frac{1}{3\sqrt{2}}$  & $-\frac{\sqrt{2}}{3}$  & $-\frac{1}{\sqrt{2}} $ & $\frac{1}{\sqrt{6}}$ \\[6pt]
$\mathcal{A}(\Xi_c^+\rightarrow \Sigma^+ \pi^-\pi^+)$  & $\frac{1}{\sqrt{6}}$ & $-\frac{1}{3\sqrt{2}}$  & $\frac{\sqrt{2}}{3}$  & $-\frac{1}{\sqrt{2}}$   &  $\frac{1}{\sqrt{6}}$ \\[6pt]
$\mathcal{A}(\Lambda_c^+\rightarrow \Sigma^+\pi^-K^+)$ & $\frac{1}{\sqrt{6}}$  & $\frac{1}{3\sqrt{2}}$ & $-\frac{\sqrt{2}}{3}$  & $\frac{1}{\sqrt{2}}$ & $\frac{1}{\sqrt{6}}$ \\[6pt]
$\mathcal{A}(\Xi_c^+\rightarrow p K^- \pi^+)$ &  $-\frac{1}{\sqrt{6}}$  & $-\frac{1}{3\sqrt{2}} $ & $\frac{\sqrt{2}}{3}$  & $\frac{1}{\sqrt{2}}$ & $\frac{1}{\sqrt{6}}$ \\[6pt]
$\mathcal{A}(\Lambda_c^+\rightarrow p \pi^-\pi^+)$   & $0$ & $\frac{\sqrt{2}}{3} $  & $\frac{\sqrt{2}}{3} $   & 0  & $\sqrt{\frac{2}{3}}$ \\[6pt]
$\mathcal{A}(\Xi_c^+\rightarrow \Sigma^+ K^-K^+)$  & $0$ & $-\frac{\sqrt{2}}{3}$  & $-\frac{\sqrt{2}}{3}$  & 0  &  $\sqrt{\frac{2}{3}}$ \\[6pt]\hline\hline
\end{tabular}
\caption{SCS decays connected to $\Lambda_c\rightarrow p\pi^+\pi^-$ by U-spin and their group-theoretical decomposition.\label{tab:group}}
\end{center}
\end{table}

\begin{table}[h]
\begin{center}
\begin{tabular}{c|c|c|c|c|c}
\hline \hline
Decay ampl. {$\mathcal{A}$} & $\Sigma (T + C_1 - E_2)$  & $\Sigma (C_2 + E_2)$ &  $\Sigma (E_1 + E_2)$ &  $\Delta (T+C_1+E_2+P_1)$  &  $\Delta (C_2+E_1+P_2)$  \\\hline\hline
$\mathcal{A}(\Lambda_c^+\rightarrow p K^-K^+)$   	& $-1$    &0  &0  &$-1$  &0    	 \\
$\mathcal{A}(\Xi_c^+\rightarrow \Sigma^+ \pi^-\pi^+)$  & $1$   &0  &0  &$-1$  & 0   \\
$\mathcal{A}(\Lambda_c^+\rightarrow \Sigma^+\pi^-K^+)$ & 0  &$-1$  &$1$  &0  & $-1$   \\
$\mathcal{A}(\Xi_c^+\rightarrow p K^-\pi^+)$  & 0  &  $1$  & $-1$  & 0  & $-1$ \\   
$\mathcal{A}(\Lambda_c^+\rightarrow p \pi^-\pi^+)$   	& $1$  &$1$ &$1$  &$-1$  & $-1$   	\\
$\mathcal{A}(\Xi_c^+\rightarrow \Sigma^+ K^-K^+)$  	& $-1$   &$-1$  &$-1$  &$-1$  &  $-1$  	\\\hline\hline
\end{tabular}
\caption{SCS decays connected to $\Lambda_c\rightarrow p\pi^+\pi^-$ by U-spin and their diagrammatical decomposition.\label{tab:topo}}
\end{center}
\end{table}

\begin{table}[h]
\begin{center}
\begin{tabular}{c|c|c|c}
\hline \hline
Decay ampl. {$\mathcal{A}_X$} & 
$\Sigma \bra{\frac{3}{2}}    2\ket{\frac{1}{2}}$   &  
$\Sigma \bra{\frac{1}{2}}_0  0\ket{\frac{1}{2}}$ &  
$\Sigma \bra{\frac{1}{2}}_1  0\ket{\frac{1}{2}}$  \\[6pt]\hline\hline
$\mathcal{A}_X(\Lambda_c^+\rightarrow p K^-K^+)$   	 & $-\frac{2}{3\sqrt{5}}$  & $\frac{1}{\sqrt{6}} $  & $-\frac{1}{3\sqrt{2}}$ \\[6pt]
$\mathcal{A}_X(\Xi_c^+\rightarrow \Sigma^+ \pi^-\pi^+)$  & $-\frac{2}{3\sqrt{5}}$  & $\frac{1}{\sqrt{6}} $  & $-\frac{1}{3\sqrt{2}}$ \\[6pt]
$\mathcal{A}_X(\Lambda_c^+\rightarrow \Sigma^+\pi^-K^+)$ & $-\frac{2}{3\sqrt{5}}$  & $-\frac{1}{\sqrt{6}} $  & $-\frac{1}{3\sqrt{2}}$ \\[6pt]
$\mathcal{A}_X(\Xi_c^+\rightarrow p K^- \pi^+)$ 	 & $-\frac{2}{3\sqrt{5}}$   & $-\frac{1}{\sqrt{6}}$  & $-\frac{1}{3\sqrt{2}} $ \\[6pt]
$\mathcal{A}_X(\Lambda_c^+\rightarrow p \pi^-\pi^+)$     & $\frac{2}{3\sqrt{5}}$  & $0$  & $-\frac{\sqrt{2}}{3}$ \\[6pt]
$\mathcal{A}_X(\Xi_c^+\rightarrow \Sigma^+ K^-K^+)$      & $\frac{2}{3\sqrt{5}}$  & $0$  & $-\frac{\sqrt{2}}{3}$ \\[6pt]\hline\hline
\end{tabular}
\caption{Decomposition of the CKM-leading U-spin breaking part of the SCS decays which are connected to $\Lambda_c\rightarrow p\pi^+\pi^-$ by U-spin. 
\label{tab:groupbreaking}}
\end{center}
\end{table}

\clearpage

\bibliography{charmbaryons.bib}
\bibliographystyle{apsrev4-1}

\end{document}